\documentclass[12pt]{article}
\usepackage{amssymb,latexsym}
\usepackage[dvips]{graphicx}
\usepackage{subeqnarray}
\newtheorem{prop}{Proposition}
\newtheorem{conc}{Conclusion}
\newtheorem{obse}{Observation}
\def\beq{\begin{equation}}
\def\eeq{\end{equation}}
\def\bea{\begin{eqnarray}}
\def\eea{\end{eqnarray}}
\def\beas{\begin{eqnarray*}}
\def\eeas{\end{eqnarray*}}
\def\nn{\nonumber}

\def\[{[\![}
\def\]{]\!]}

\def\a{\alpha}

\def\o{\omega}
\def\t{\theta}
\def\T{\Theta}
\def\r{{\hat r}}
\def\p{{\hat p}}
\def\R{{\hat R}}
\def\P{{\hat P}}
\def\F{{\hat F}}
\def\G{{\hat G}}
\def\hP{{\hat P}}
\def\hR{{\hat R}}
\def\hp{{\hat p}}
\def\hM{{\hat M}}
\def\hr{{\hat r}}

\def\P{{\cal P}}

\def\hbP{{\hat {\bf P}}}
\def\hbR{{\hat {\bf R}}}
\def\hbM{{\hat {\bf M}}}
\def\hbp{{\hat {\bf p}}}
\def\hbr{{\hat {\bf r}}}

\def\ra{\rangle}
\def\la{\langle}

\def\-{$\bullet\;$}
\def\e{\varepsilon}

\def\dag{\dagger}

\def\|{\,|\,}
\def\br{\bar r}
\def\bp{\bar p}

\def\ZZ{\mathbb{Z}}

\renewcommand{\theequation}{\arabic{section}.\arabic{equation}}
\begin{document}
%
%
\begin{center}
{\Large \bf
The non-commutative and discrete spatial structure of a 3D Wigner
quantum oscillator}\\[5mm]
{\bf R.C.~King}\footnote{E-mail: R.C.King@maths.soton.ac.uk}\\[1mm]
Faculty of Mathematical Studies, University of Southampton,\\
Southampton SO17 1BJ, U.K.;\\[2mm]
{\bf T.D.\ Palev}\footnote{E-mail: tpalev@inrne.bas.bg.
Permanent address:
Institute for Nuclear Research and Nuclear Energy,
Boul.\ Tsarigradsko Chaussee 72, 1784 Sofia, Bulgaria.}\\[1mm]
Abdus Salam International Centre for Theoretical Physics,\\
PO Box 586, 34100 Trieste, Italy;\\[2mm]
{\bf N.I.~Stoilova}~\footnote{E-mail: Neli.Stoilova@rug.ac.be.
Permanent address:
Institute for Nuclear Research and Nuclear Energy,
Boul.\ Tsarigradsko Chaussee 72, 1784 Sofia, Bulgaria.}
{\bf and J.\ Van der Jeugt}\footnote{E-mail:
Joris.VanderJeugt@rug.ac.be.}\\[1mm]
Department of Applied Mathematics and Computer Science,
University of Ghent,\\
Krijgslaan 281-S9, B-9000 Gent, Belgium.
\end{center}

\vskip 10mm
\begin{abstract}
The properties of a non-canonical 3D Wigner quantum
oscillator, whose position and momentum operators generate
the Lie superalgebra $sl(1|3)$, are further investigated.
Within each
state space $W(p)$, $p=1,2,\ldots$, the energy $E_q$, $q=0,1,2,3$,
takes no more than 4 different values. If the oscillator is in a
stationary state $\psi_q\in W(p)$ then measurements of the
non-commuting Cartesian coordinates of the particle are such
that their allowed values are consistent with it being found
at a finite number of sites, called ``nests''. These lie on a
sphere centered on the origin of fixed, finite radius $\varrho_q$.
The nests themselves are at the vertices of a rectangular
parallelepiped.
In the typical cases ($p>2$) the number of nests is 8
for $q=0$ and $3$, and varies from 8 to 24, depending on the state,
for $q=1$ and $2$.
The number of nests is less in the atypical cases ($p=1,2$),
but it is never less than two. In certain states in $W(2)$
(resp. in $W(1)$) the oscillator is ``polarized'' so that all
the nests lie on
a plane (resp. on a line). The particle cannot be
localized in any one of the available nests alone since the
coordinates do not commute. The probabilities of measuring
particular values of the coordinates are discussed.
The mean trajectories and the standard deviations of the coordinates
and momenta are computed, and conclusions are drawn about
uncertainty relations.
\end{abstract}
\vfill\eject

\section{Introduction}
\setcounter{equation}{0}

In the present paper we continue the investigation of new quantum systems
originating from the representation theory of basic classical Lie
superalgebras.
In particular, we study further the properties of a 3-dimensional (3D)
Wigner quantum  oscillator whose
mathematical background involves the Lie superalgebra
$sl(1|3)$~\cite{1,2}.

The idea itself behind these investigations stems from the 1950's paper
of Wigner
{\em Do the equations of motion determine the quantum mechanical commutation
relations?}~\cite{3}.
In this paper Wigner has generalized a result of Ehrenfest~\cite{4}.
The latter stated (up to ordering details, which are irrelevant in our case)
that in the Heisenberg picture of quantum mechanics
Hamilton's (resp.\ the Heisenberg)
equations are a unique consequence of the canonical commutation relations
(CCRs) and
the Heisenberg (resp.\ Hamilton's) equations. Wigner has proved a stronger
statement.
He has shown through an example that Hamilton's equations can be identical
to the
Heisenberg equations even if the position and momentum operators do not
satisfy the CCRs.

Wigner's example was a one-dimensional oscillator with a Hamiltonian
${\hat H}={1\over 2}(\hp^2+{\hat q}^2)$, in units such that $m=\omega=\hbar=1$. 
Abandoning the requirement $[{\hat q},\hp]=i$, Wigner
searched for all operators $\hat q$ and $\hp$ such that Hamilton's equations ${\dot
{\hat q}}=\hp$ and ${\dot \hp}=-{\hat q}$ are identical with the Heisenberg equations
${\dot {\hat q}}=i[{\hat H},{\hat q}]$ and ${\dot \hp}=i[{\hat H}, \hp]$. 
In addition to the canonical solution he found infinitely many other solutions,
that is infinitely many solutions for the pair ${\hat q}$ and ${\hat p}$.
He interpreted these as position and momentum operators
despite the fact that they do not satisfy the CCRs $[{\hat q},\hp]=i$.

Different aspects of Wigner's idea were studied by several authors. Among the earlier
references we mention~\cite{5}-\cite{10b}, but the subject still remains of
interest~\cite{11}-\cite{16c}.

The key motivation for a generalization of the concept of a quantum system ~\cite{1,2} comes
from the observation of Wigner ~\cite{3} that the Heisenberg equations and 
Hamilton's equations have a more immediate physical significance than the canonical
commutation relations. From this point of view it is logically justified to
postulate as the starting point the Heisenberg equations and Hamilton's equations instead
of the CCRs.

The conjecture that the CCRs have to be modified, including the possibility that the
configuration space coordinates may not mutually commute, originated recently  from
string theory (we refer to~\cite{ga95} for a survey on the subject) and also from
quantum groups (see~\cite{CDP96} for a review). However, the idea itself was already suggested by
Heisenberg in the late 1930's (as explained in~\cite{RSW}), and perhaps the first example
of this kind was given by Snyder~\cite{Sn47}.

The above observations justify the consideration of what we call a Wigner quantum
system~\cite{1,2}. As in~\cite{2}, a Wigner quantum system (WQS) differs from a
canonical quantum system only by the replacement of the postulate on canonical
commutation relations by a new postulate. In particular, consider an $n$-particle
system in three dimensions with Hamiltonian 
\beq
      {\hat H} =
\sum_{\a=1}^n {{{ \hbP}_\a^2}\over{2m_\a}} + V({ \hbR}_1,{ \hbR}_2,\ldots,{ \hbR}_n),
\label{1.1} 
\eeq 
which depends on the $6n$ variables ${\hbR}_\a$ and ${\hbP}_\a$, with
$\a=1,2,\ldots,n$, to be interpreted as (Cartesian) coordinates and momenta,
respectively. Just as in ordinary quantum mechanics, the following conditions should
hold:
\begin{itemize}
\item[{\bf P1}] The state space
 $W$ is a Hilbert space. To every physical
observable ${\cal O}$ there corresponds a Hermitian (self-adjoint) operator $\hat O$
acting in $W$.
\item[{\bf P2}] The observable ${\cal O}$ can take on only those values which
are eigenvalues of $\hat O$. The expectation value of the observable ${\cal O}$ in a
state $\psi$ is given by $\la {\hat O}\ra_\psi=(\psi,{\hat O}\psi)/(\psi,\psi)$, where
$(\psi,\phi)$ denotes the scalar product of $\psi,\phi\in W$.
\end{itemize}
Together with others, these postulates are common to any quantum system. We repeat only
{\bf P1} and {\bf P2} here because we will explicitly make use of them. The difference
with canonical quantum mechanics comes from postulating {\bf P3}:
\begin{itemize}
\item[{\bf P3}] Hamilton's equations and the Heisenberg equations hold and
are identical (as operator equations) in $W$.
\end{itemize}
In the canonical case instead of {\bf P3} one postulates the validity of the Heisenberg
equations and the CCRs. Then, as mentioned above, Hamilton's equations hold too.

Here, we shall no longer use the CCRs, but rely on {\bf P3} instead. The corresponding
system is called a Wigner quantum system. Although ${\hbR}_\a$ and ${\hbP}_\a$ no
longer satisfy the CCRs, following Wigner (and the related papers ~\cite{5}-\cite{16c},
\cite{Sn47}),
we shall still interpret them as the operators corresponding to 
measurements of the physical position
and momentum of the WQS and refer to them as position and momentum operators.

Let us point out again that the above postulates do not provide a complete description
of Wigner quantum systems nor of Wigner quantum oscillators to be
studied in the present paper. On the ground of these postulates alone one cannot
determine the expressions for the operators of the angular momentum, for instance.

It is perhaps worth mentioning that postulate {\bf P3} can be
weakened (so far only in the 1D case) in a manner consistent
with Wigner's ideas~\cite{10b}
so that deformed quantum oscillators~\cite{Bie,Mac} and, more
generally, Daskaloyannis oscillators~\cite{10b,Das} can be viewed
as (generalized) Wigner oscillators.

Our approach to Wigner quantum oscillators is essentially based
on two observations. The first one, due to  Kamefuchi and
Ohnuki~\cite{17}, is the  proof that  all  solutions found by Wigner are
different representations of {\em just
one pair} of para-Bose (pB) creation and annihilation operators
(CAOs) $B^\pm=(q\mp p)/\sqrt2$.
More generally, we recall that the
(representation independent) pB operators, which generalize
Bose statistics, are defined  by the relations~\cite{19}:
\beq
[\{B_i^\xi,B_j^\eta\},B_k^\zeta]=(\zeta-\xi)\delta_{ik}B_j^\eta
+(\zeta-\eta)\delta_{jk}B_i^\xi,
\ \ i,j,k=1,2,\ldots,N,\ \ \xi,\eta,\zeta=\pm,
\label{1.2}
\eeq
(where, by convention, $\xi$, $\eta$, $\zeta$ are written as
$\pm$ when used as superscripts, and as $\pm 1$ when used algebraically
in the factors $(\zeta-\xi)$ and $(\zeta-\eta)$).
Here and throughout the paper $\{x,y\}=xy+yx$ and $[x,y]=xy-yx$
for any $x,y$.

The second relevant  observation is that any $N$ pairs of pB
operators $B_1^\pm, \ldots, B_N^\pm$ are odd elements, generating
a Lie superalgebra~\cite{20}, isomorphic to the orthosymplectic Lie
superalgebra $osp(1|2N)$~\cite{18}. The Fock spaces of any $N$ pairs
of parabosons and in particular of bosons are irreducible
$osp(1|2N)$ modules. In this terminology the oscillator of Wigner
can be called an $osp(1|2)$ oscillator, its position and momentum
operators are the odd generators of $osp(1|2)$, the Hamiltonian
is a simple polynomial of the position and momentum operators and
the solutions found by Wigner are different irreducible
representations of this Lie superalgebra.

The $osp(1|2N)$ Lie superalgebra is a basic classical Lie
superalgebra from class $B$ in the classification of Kac~\cite{21}. In
fact~(\ref{1.2}) yields one possible definition of $osp(1|2N)$: the
associative superalgebra
with unity, subject to the relations~(\ref{1.2}) is
the universal enveloping algebra of $osp(1|2N)$.

The results of Wigner can be easily extended to any $N$-dimensional
harmonic oscillator, turning it into a Wigner quantum oscillator (WQO).
To this end one has first to express the Hamiltonian via Bose
operators
\beq
    [b_{i}^-,b_{j}^+]=\delta_{ij}
     \quad  [b_{i}^+,b_{j}^+]=[b_{i}^-,b_{j}^-]=0,
  \  \ \  i,j=1,\ldots,N,
\label{1.3}
\eeq
and their anticommutators and subsequently replace them
with pB operators~(\ref{1.2}). The
corresponding solutions are now associated with
infinite-dimensional irreducible representation of the Lie superalgebra
$osp(1|2N)$. Since this superalgebra is of class $B$ we refer to the
statistics of the canonical quantum oscillator
(CQO) (and its pB generalization) as being $B$-superstatistics.

Having observed all this, it was natural to ask whether one can satisfy
the postulates {\bf P1-P3} with position and momentum operators which generate
algebras from the other, different from $B$,
classes of basic Lie superalgebras.
A positive answer to this question was given in~\cite{1,2} with
operators $A_{i}^\pm, ~i=1,\ldots,N,$ which satisfy
certain relations that we will specify in the next section.
These ensure that they generate the Lie superalgebra $sl(1|N)$.
The corresponding solutions are this time associated with a
Wigner quantum system  that takes the form of an $N$-dimensional
{\em non-canonical} Wigner quantum oscillator.
Since the special linear
Lie superalgebra $sl(1|N)$ is of class $A$ we refer to the statistics
of this WQO as being $A$-superstatistics.

In a similar approach Barut and Bracken~\cite{22} have described
the internal dynamics ({\em Zitterbewegung}) of Dirac's electron.
Their creation and annihilation operators
satisfy similar triple relations as in our
case (Eqs.~(\ref{2.11})),
but instead of
a Lie superalgebra they generate the Lie algebra $so(5)$.

In the present paper we study further the properties of the
3D WQO, related to the $sl(1|3)$ superalgebra and initiated in~\cite{1,2}.
The paper is organized as follows.

In Section~2 we outline the mathematical structure of the
3D non-canonical oscillator. The compatibility
between the Heisenberg equations and the Hamilton's equations
is achieved with operators $A_1^\pm, A_2^\pm,A_3^\pm$
which satisfy triple
relations similar to those for the para-Bose case~(\ref{1.2}),
but this time they generate the Lie superalgebra
$sl(1|3)$. The Fock spaces $W(p)$ of these operators are
defined. The inequivalent representations are
labeled by one positive integer~$p$.  For $p>2$
all Fock spaces are 8-dimensional, whereas in the case
$p=1$ (resp.\ $p=2$) $\dim \ W(1)=4$ (resp.\ $\dim\ W(2)=7$).
In the terminology of Kac~\cite{21} the $p=1,2$
representations are called
atypical representations. For this reason we refer to the
Fock spaces $W(1)$ and $W(2)$ as to atypical, to the
corresponding oscillator as to atypical etc. We
shall see in the next sections that the properties of the
atypical oscillators are very different from those
with $p>2$.

In Section~3 we recall the known~\cite{1,2} physical properties of the $sl(1|3)$ WQOs.
Firstly, the oscillator has finite space dimensions and the Hamiltonian has no more
than 4 different eigenvalues. In the stationary states the distance of the particle to
the origin is quantized so that the particle is constrained to move on  one of  4
possible spheres. Secondly, the geometry of the oscillator is non-commutative. 
The various coordinates do not commute with one another, nor do the various components
of the momentum. Therefore, in particular, the position of the
particle on the corresponding sphere cannot be localized. In this respect the WQO
belongs to the class of models of non-commutative quantum oscillators~\cite{P1}-\cite{D}
and, more generally, to theories with non-commutative geometry~\cite{E,F}.
It is shown, however, that the non-commutativity between our position and
momentum operators is different from the non-commutativity appearing in
the most commonly adopted form of  generalized Heisenberg
commutation relations (see eq.~(\ref{3.4a})).

All results after Section~3 are new. In Section~4 the
probabilistic distribution of the particle is analyzed. The basis
consists of stationary states. The main result is the following:
if the particle is in one of the basis states with $p>2$, then
measurements of its coordinates are consistent with it only
being found at 8 particular points on
a sphere which form
the vertices of a rectangular parallelepiped (see
Figure~1). Thus as in~\cite{23}, the coordinates of the
particle are observables with a quantized spectrum just like
energy, angular momentum, etc. The number of points, called
``nests'', can be even less in the atypical cases. In certain
states with $p=2$ the oscillator  is becoming a flat object with
4 vertices (Figure~2). There are three states in the $p=1$ case
when the oscillator is even one-dimensional (Figure~3).
In Section~5 the mean trajectories and the standard deviations of the position
and  momenta operators for an arbitrary state are written down.
It is shown (Conclusion~3) that there exists no nontrivial
analogue of the Heisenberg uncertainty relations since one can
always find a state $x$ for which either $(\Delta r_k)_x=0$ or
$(\Delta p_k)_x=0$.

In Section~6 we show that despite the fact
that the $sl(1|3)$ oscillator is very different from the 3D
canonical oscillator, they still have some features in common. In
particular we show that to each $p=1$ mean trajectory of the
$sl(1|3)$ oscillator there corresponds exactly the same
trajectory of the 3-dimensional canonical oscillator.

\section{Mathematical structure of the 3D WQO}
\setcounter{equation}{0}

Let $\hat H$ be the Hamiltonian of a  three-dimensional
harmonic oscillator, that is
\beq
{\hat H}={{\hbP}^2 \over 2m}
+{m\omega^2\over 2}{\hbR}^2.
\label{2.1}
\eeq
We proceed to view this oscillator as a Wigner quantum system and
work throughout in the Heisenberg picture in which the operators
are, in general, time dependent. According to postulate {\bf P3}
the operators $\hbR$ and $\hbP$ have to be defined in such a way
that Hamilton's equations
\beq
    {\dot \hbP}=-m\omega^2 \hbR, \ \ {\dot \hbR} = {1\over m}\hbP
\label{2.2}
\eeq
and the Heisenberg equations
\beq
     {\dot \hbP} = {i\over{\hbar}}[{\hat H},\hbP], \ \
     {\dot \hbR} = {i\over{\hbar}}[{\hat H},\hbR]
\label{2.3}
\eeq
are both valid, and are identical as operator equations. These
equations are compatible only if
\beq
   [{\hat H},\hbP]=i\hbar m \omega^2 \hbR ,\ \
   [{\hat H},\hbR]=-{{i\hbar}\over{m}} \hbP.
\label{2.4}
\eeq

The most general solution of~(\ref{2.2}) and~(\ref{2.3})
is not known. Here
we mention  the canonical Bose solution. Expressed via boson
creation and annihilation operators it reads:
\beq
r_k(t)=\sqrt{\hbar\over{2m\omega}}~(b_k^+e^{~i \omega t}
+ b_k^-~e^{-i \omega t} ),
\quad p_k(t)=i\ \sqrt{m\omega\hbar\over 2}~
(b_k^+e^{~i \omega t}
-~ b_k^-~e^{-i \omega t} ).
\label{2.5}
\eeq
In this setting $r_k$ and $p_k$ are canonical position and momentum
operators, defined in a Bose Fock space $\Phi$ with  orthonormal
basis states
\beq
|n_1,n_2,n_3)={(b_1^+)^{n_1} (b_2^+)^{n_2}~ (b_3^+)^{n_3}
\over{\sqrt{n_1!~ n_2!~ n_3!}}}|0\ra,\quad
n_1,~ n_2,~ n_3 \in \ZZ_+
\label{2.6}
\eeq
subject to the known transformation relations:
\beq
b_k^+ |\ldots,n_k,\ldots)=\sqrt{n_k+1}|\ldots,n_k+1,\ldots),\quad
b_k^- |\ldots,n_k,\ldots)=\sqrt{n_k}|\ldots,n_k-1,\ldots).
\label{2.7}
\eeq
As mentioned already in the Introduction, this Bose solution
belongs to the class of $B$-superstatistics.

In the present paper we deal with solutions of~(\ref{2.2}) and~(\ref{2.3})
for which the operators $\hR_i$ and $\hP_i$, $i=1,2,3,$  generate a Lie
superalgebra from the class $A$, more precisely $sl(1|3)$.
{\em We emphasize again that despite the fact that $\hR_i$ and $\hP_i$
do not satisfy the CCRs, i.e.\ they are  only
``position-like'' and ``momentum-like''
operators, we interpret them as  operators describing position
and momentum.}
Accordingly we refer to them as position and momentum
operators in the rest of this paper.
To make the connection with $sl(1|3)$ we write the operators
$\hbP\equiv(\hP_1,\hP_2,\hP_3)$ and
$\hbR\equiv(\hR_1,\hR_2,\hR_3)$ in terms of new operators:
\beq
A_{k}^\pm=\sqrt{m \omega \over 2 \hbar} \hR_{k} \pm
{i  \sqrt {1\over 2m \omega \hbar}}\hP_{k}, \qquad k=1,2,3.
\label{2.8}
\eeq
The Hamiltonian $\hat H$ of~(\ref{2.1}) and the compatibility
conditions~(\ref{2.4})  then take the form:
\beq
    {\hat H} = {{\omega\hbar}\over{2}}\sum_{i=1}^3 \{A_{i}^+,A_{i}^-\},
\label{2.9}
\eeq
\beq
\sum_{i=1}^3  [ \{A_{i}^+,A_{i}^- \},A_{k}^\pm] =\mp 2A_{k}^\pm ,
\quad i,k=1,2,3.
\label{2.10}
\eeq
As a solution to~(\ref{2.10}) we chose operators $A_{i}^\pm$ that
satisfy the
following triple relations:
\begin{subeqnarray}
&& [\{A_{i}^+,A_{j}^-\},A_{k}^+]=
\delta_{jk}A_{i}^+ -\delta_{ij}A_{k}^+, \slabel{2.11a}\\
&& [\{A_{i}^+,A_{j}^-\},A_{k}^-]=
-\delta_{ik}A_{j}^- +\delta_{ij}A_{k}^-, \slabel{2.11b}\\
&& \{A_{i}^+,A_{j}^+\}= \{A_{i}^-,A_{j}^-\}=0. \slabel{2.11c}
\label{2.11}
\end{subeqnarray}
In our case $i,j,k=1,2,3$. Equations~(\ref{2.11})
are defined however for $i,j,k=m,m+1,\ldots, n$, where
$m$ and $n$ are any integers
(including $m=-\infty$ and $n=\infty$).

\begin{prop}
The operators $A_{i}^\pm$, $i=1,\ldots,n$,
satisfying~(\ref{2.11}),
are odd elements generating the Lie superalgebra $sl(1|n)$~\cite{24}.
\end{prop}

The generators $A_i^\pm$, $i=1,\ldots,n$
are said to be
creation and annihilation operators
of $sl(1|n)$. They resemble ordinary Fermi operators (see~(\ref{2.11c}))
and can be interpreted as quasi-fermions in the context of generalized
statistics~\cite{PSJ}.
These CAOs are the analogue of the Jacobson generators for the Lie
algebra $sl(n+1)$~\cite{25} and could also be called Jacobson
generators of $sl(1|n)$.

Coming back to the 3D oscillator, we emphasize again that all
considerations here are in the Heisenberg picture. The position
and momentum operators depend on time. Hence also the CAOs depend
on time. Writing this time dependence explicitly, one has:
\bea
& \hbox{Hamilton's equations} & {\dot
A}_{k}^\pm(t)=\mp i \omega A_{k}^\pm(t), \label{2.12} \\
& \hbox{Heisenberg equations} & {\dot A}_{k}^\pm(t)={{i \omega
}\over{2}}\sum_{i=1}^3 [\{A_{i}^+(t), A_{i}^-(t)\},A_{k}^\pm(t)].
\label{2.13}
\eea
The solution of~(\ref{2.12}) is evident,
\beq
A_{k}^\pm(t)=\exp(\mp i \omega t) A_{k}^\pm(0)
\label{2.14}
\eeq
and therefore if the defining relations~(\ref{2.11}) hold
at a certain time $t=0$, i.e., for $A_{k}^\pm \equiv A_{k}^\pm(0) $,
then they hold as equal time relations for any other time $t$.
{}From~(\ref{2.11})
it follows also that the equations~(\ref{2.12})
are identical with equations~(\ref{2.13}). For
further use we write the time dependence of
$\hbR=(\hR_1,\hR_2,\hR_3)$
and $\hbP=(\hP_1,\hP_2,\hP_3)$ explicitly:
\begin{subeqnarray}
&& \hR_{k}(t)={\sqrt{\hbar \over {2m\omega} }} (A_{k}^+
e^{-i \omega t} +A_{k}^-e^{i \omega t}), \slabel{2.15a} \\
&& \hP_{k}(t)=-i\sqrt{m \omega \hbar \over {2}}
(A_{k}^+ e^{-i \omega t}- A_{k}^-e^{i \omega t}), \slabel{2.15b}
\label{2.15}
\end{subeqnarray}
where $k=1,2,3$.

Next, we wish to introduce the concept of angular momentum and the
related space rotations.
In order to define the angular momentum operator $\hbM = (\hM_1, \hM_2, \hM_3)$,
we assume (as in canonical quantum mechanics, for observables having a 
classical analogue) that its components are in the enveloping algebra 
of $\hbR=(\hR_1,\hR_2,\hR_3)$ and $\hbP=(\hP_1,\hP_2,\hP_3)$
and moreover that they are linear with respect the components of these
operators. We also require that $\hM_1$, $\hM_2$ and $\hM_3$ commute with the 
Hamiltonian, and that they span a basis of the Lie algebra $so(3)$. 
The operators with the required properties are:
\beq
      \hM_{j} = -{{1}\over{\hbar}} \sum_{k,l=1}^3
    \ \epsilon_{jkl} \{\hR_{k}, \hP_{l}\},
      \qquad j=1,2,3,
\label{2.16}
\eeq
($\epsilon_{jkl}$ is the antisymmetric unit tensor of rank~3), which
take the following form in terms of the CAOs~(\ref{2.11})
\beq
      \hM_{j} = -i \sum_{k,l=1}^3
    \ \epsilon_{jkl} \{A_{k}^+, A_l^-\},
      \qquad j=1,2,3.
\label{2.17}
\eeq
It is straightforward to verify that these operators satisfy the 
required commutation relations:
\beq 
[\hM_j,\hM_k]=i\epsilon_{jkl}\hM_l,
\qquad 
[\hM_j, {\hat H} ]=0. \label{so} 
\eeq
Hence $\hM_1$, $\hM_2$, $\hM_3$ are the generators of the Lie algebra $so(3)$, 
i.e.\ they generate the $so(3)$ subalgebra of $sl(1|3)$.
Moreover they are integrals of motion: they do not depend on time.
We shall interpret $\hbM$ as the operator describing the angular momentum
of the WQO, and simply refer to it as the angular momentum.
As in the canonical case, we shall also identify the components of
$\hbM$ as the operators of infinitesimal rotations. Then the relations
\beq
[\hM_j,\hR_{ k}]=i\epsilon_{jkl}\hR_{ l}, 
\qquad [\hM_j,\hP_{k}]=i\epsilon_{jkl}\hP_{ l},
\eeq
together with~(\ref{so}), show that $\hbR$, $\hbP$ and $\hbM$ all
transform as vector operators with respect to space rotations.

The state spaces which we consider here are those irreducible
$sl(1|3)$ modules that may be
constructed  by means of the usual
Fock space technique precisely as in the parastatistics case~\cite{19}.
To this end we require that the
representation space, $W(p)$, contains
(up to a multiple) a unique cyclic vector $|0\ra$ such that
\beq
    A_{i}^-\|0\ra=0, \quad
    A_{i}^-A_{j}^+\|0\ra=p\delta_{ij}\|0\ra,
\quad i,j=1,2,3.
\label{2.18}
\eeq

The above relations are enough for the construction of the full
representation space $W(p)$. This space defines an
indecomposable finite-dimensional
representation of the CAOs~(\ref{2.11}) and hence of
$sl(1|3)$ for any value of~$p$. However, following {\bf P1}
and {\bf P2}, we wish to impose the
further physical requirements that:
\begin{itemize}
\item[(a)]  $W(p)$ is a Hilbert space with respect to the natural
Fock space inner product;
\item[(b)] the observables, in particular the position and momentum
operators~(\ref{2.15}), are Hermitian operators.
\end{itemize}
Condition (b) reduces to the requirement that the Hermitian conjugate of
$A_{i}^+$ should be $A_{i}^-$, i.e.
\beq
      (A_{i}^\pm)^\dag=A_{i}^\mp.
\label{2.19}
\eeq
The condition (a) is then such that $p$ is restricted to be a
positive integer~\cite{24}, in fact any positive integer.

Let $\T \equiv(\t_{1},\t_{2},\t_{3})$.
The state space $W(p)$ of the system is spanned by the following
orthonormal basis (called the $\Theta$-basis):
\beq
|p;\T\ra\equiv|p;\t_{1},\t_{2},\t_{3}\ra =
{\sqrt{{(p-q)!}\over{p!}}}\
(A_{1}^+)^{\t_{1}} (A_{2}^+)^{\t_{2}} (A_{3}^+)^{\t_{3}} \ |0\ra,
\label{2.20}
\eeq
where
\beq
 \t_{i}\in\{0,1\}\ \ \hbox{for all}\ \  i=1,2,3
\label{2.21}
\eeq
and
\beq
0 \leq q \equiv \t_{1} + \t_{2} + \t_{3} \leq \min(p,3).
\label{2.22}
\eeq
The transformation of the basis states~(\ref{2.20}) under the action of
the CAOs reads as follows:
\begin{subeqnarray}
&& A_{i}^-\|p;\ldots,\t_i,\ldots\ra
=\t_{i}(-1)^{\t_1+\cdots +\t_{i-1}}\sqrt{p-q +1}\
|p;\ldots,\t_i-1,\ldots\ra, \slabel{2.23a}\\[2mm]
&&  A_{i}^+\|p;\ldots,\t_i,\ldots\ra
=(1-\t_{i})(-1)^{\t_1+\cdots +\t_{i-1}}\sqrt{p-q}\
|p;\ldots,\t_i+1,\ldots\ra. \slabel{2.23b}
\label{2.23}
\end{subeqnarray}
The factors $\t_{i}$ and $(1-\t_{ i})$ ensure that
the only non-vanishing cases are those for which
$|p;\ldots,\t_i\pm 1,\ldots\ra$
do indeed belong to the set of basis states
defined by~(\ref{2.20})-(\ref{2.22}).

Note the first big difference between this non-canonical WQO and
the case of a conventional CQO:
\begin{obse}
Contrary to the CQO with an
infinite-dimensional state space,  each state space $W(p)$ of the
WQO is finite-dimensional.
\end{obse}
In fact $\dim W(p)=8$ for $p>2$, whereas $\dim W(1)=4$ and $\dim
W(2)=7$.

\section{Known properties of 3D WQOs}
\setcounter{equation}{0}

Here we recall the physical properties of the
Wigner quantum oscillators  as given in~\cite{1,2}.

The first thing we note is that the representation of
$sl(1|3)$ was chosen such that,
as in the case of a 3D CQO, the
physical observables $\hat H$, $\hbR$, $\hbP$ and  $\hbM$
are, in the case of the WQO, all
Hermitian operators within every Hilbert space $W(p)$ for each
$p=0,1,\ldots$ (in accordance with postulate {\bf P1}).

Secondly, in the case of the WQO the Hamiltonian $\hat H$ is diagonal
in the basis~(\ref{2.20})-(\ref{2.22}),
i.e.\ the basis vectors $|p;\T\ra$ are
stationary states of the system. As in  the 3D CQO the energy
levels are equally spaced  with the same spacing $\hbar\omega$.
Contrary to the CQO each Hilbert space $W(p)$ has no more
than four  equally spaced energy levels, with
spacing $\hbar\omega$. More precisely,
\beq
     {\hat H}\|p;\T\ra = E_q\|p;\T\ra\ \ \hbox{with}\ \
     E_q={\hbar\omega\over 2}\left(3p-2q \right).
\label{3.1}
\eeq
So we can define stationary states $\psi_q$ as superpositions of
states $|p;\T\ra$ with the same $q$:
\begin{subeqnarray}
&& \psi_0 = |p;0,0,0\ra, \slabel{3.2a} \\
&& \psi_1 = \alpha(1,0,0)|p;1,0,0\ra + \alpha(0,1,0)|p;0,1,0\ra +
\alpha(0,0,1)|p;0,0,1\ra , \slabel{3.2b} \\
&& \psi_2 = \alpha(1,1,0)|p;1,1,0\ra + \alpha(1,0,1)|p;1,0,1\ra +
\alpha(0,1,1)|p;0,1,1\ra , \slabel{3.2c} \\
&& \psi_3 = |p;1,1,1\ra, \slabel{3.2d}
\label{3.2}
\end{subeqnarray}
where $\alpha(\t_1,\t_2,\t_3)$ are complex numbers. The
stationary states satisfy ${\hat H}\psi_q = E_q \psi_q$. Only the
states with $q\leq p$ belong to the space $W(p)$. Note that in
the atypical cases ($p=1,2$) the lowest energy level is
degenerate: there are three linearly independent states with the
same ground state energy.

Perhaps  the most striking difference between  the WQO and the
CQO is that the geometry of the Wigner oscillators is
non-commutative: the position operators $\hR_1$, $\hR_2$, $\hR_3$ of
the oscillating particle  do not commute with each other,
\beq
[\hR_i,\hR_j]\ne 0 \quad \hbox{ for } \quad i\ne j=1,2,3.
\label{3.3}
\eeq
Hence for the Wigner oscillators {\em a coordinate representation
($x$-representation) does not exist}. Similarly,
\beq
[\hP_i,\hP_j]\ne 0 \quad \hbox{ for } \quad i\ne j=1,2,3
\label{3.4}
\eeq
and therefore also {\em a momentum representation ($p$-representation)
cannot be defined}.

The relations~(\ref{3.3}) and (\ref{3.4}) imply that the WQO belongs to the class of
models of non-commutative quantum oscillators~\cite{P1}-\cite{D} 
and, more generally, to
theories with non-commutative geometry~\cite{E,F}. The literature on this subject is
vast. Moreover the subject is not any longer of purely theoretical interest. Most
recently papers predicting (experimentally) measurable deviations from the
commutativity of the coordinates have been published~\cite{G}-\cite{Jellal}. 
Here however, we only deal with a purely theoretical description.

Following the non-commutativity, it is natural to ask about the nature
and value of the commutator $[\hR_i,\hR_j]$
(or $[\hP_i,\hP_j]$, or $[\hR_i,\hP_j]$). The answer to this question is
relevant since it is used in the derivation of the uncertainty relations between the
coordinates for instance. 
To answer this, note that the commutators between the operators $\hR_i$ and $\hP_j$
do not belong to the Lie superalgebra $sl(1|3)$ 
($\hR_i$ and $\hP_i$ are odd elements of the algebra),
so they cannot be rewritten in a simpler form.
Of course, one can compute the action of these commutators on basis
vectors of the considered $sl(1|3)$ modules $W(p)$. 
We have actually made these computations. However the
resulting formulas are rather complicated. Later, we content ourselves
with just one illustrative example in (\ref{5.18}).

It is worth pointing out that the commutators $[\hR_1,\hR_2]$,
$[\hR_1,\hR_3]$ and $[\hR_2,\hR_3]$ are themselves operators that 
do not commute with each other in any one of the state
spaces $W(p)$, $p=1,2,\ldots$ and therefore cannot be diagonalized simultaneously.
For this reason the non-commutativity between our position and momentum
operators is very different from the non-commutativity of generalized Heisenberg
commutation relations
\beq
[\hr_i,\hr_j]=i\theta_{ij},\quad
[\hr_i,\hp_j]=i\hbar\delta_{ij},\quad [\hp_i,\hp_j]=i{\bar\theta_{ij}} 
\label{3.4a}
\eeq
often adopted in the literature on non-commutative quantum mechanics (see for instance
~\cite{P1}-\cite{D}). In the right hand side of~(\ref{3.4a}) $\theta_{ij}$
and ${\bar\theta_{ij}}$ are numbers (which are often further simplified, e.g.,
${\bar\theta_{ij}}=0$, ${\theta_{ij}}=\theta$, etc.) and therefore all operators
$[\hr_i,\hr_j]$, $[\hr_i,\hp_j]$, $[\hp_i,\hp_j]$ are simultaneously diagonal (in
any basis of the state space).

Turning back to the WQO, there are two interesting integrals of motion,
namely ${\hbR}^2$ and ${\hbP}^2$. Furthermore, they are proportional
to $\hat H$:
\beq
{\hat\epsilon} \equiv
{2\over \omega \hbar} {\hat H} = { 2 m \omega \over \hbar} {\hbR}^2
= {2 \over m\omega\hbar} {\hbP}^2 = \sum_{i=1}^3 \{ A_i^+,A_i^-\}.
\label{3.5}
\eeq
And thus:
\begin{subeqnarray}
&&{\hbR}^2\|p;\T\ra={\hbar\over{2m\omega }}\left(3p-2q \right)\|p;\T\ra,
     \slabel{3.6a}\\
&&{\hbP}^2\|p;\T\ra={m\omega\hbar\over{2}}\left(3p-2q \right)\|p;\T\ra,
     \slabel{3.6b}
\label{3.6}
\end{subeqnarray}
for $0 \leq q \equiv \t_{1} + \t_{2} + \t_{3} \leq \min(p,3)$.
Eq.~(\ref{3.6a}) indicates that if the oscillator is in a stationary
state $\psi_q$ with energy $E_q={\hbar\omega\over 2}\left(3p-2q
\right)$, then the distance $\varrho_q$ between the oscillating
particle and the origin of the coordinate system is
\beq
\varrho_q=\sqrt{ {\hbar\over{2m\omega }}\left(3p-2\t_1-2\t_2-2\t_3
\right)}
\label{3.7}
\eeq
and this distance is an integral of motion,
it is preserved in time. For further references we formulate the following
observation.

\begin{conc}
Each stationary state  $\psi_q$, which is
a superposition of states $|p;\T\ra$ with one and the same
$q=\t_1+\t_2+\t_3$, corresponds to a configuration in which  the
particle is somewhere at a distance  $\varrho_q$ from the centre
of the coordinate system. However, the position of the particle on
the sphere of radius $\varrho_q$ cannot be localized because the
coordinates do not commute with one another.
\end{conc}

The maximum distance of the particle from the centre is
\beq
\varrho_{max}\equiv \varrho_0=\sqrt{3\hbar p\over 2m\omega}
\label{3.8}
\eeq
and this corresponds to the state $|p;0,0,0\ra$, which carries
also the maximal energy $E_{max}={3\over 2} \hbar\omega p$. Thus
the WQO occupies a finite volume. The oscillating particle is
locked in a sphere with radius~(\ref{3.8}), which is another property
very different from the CQO for which there is no finite upper
bound on the radial distance.

Finally we note that
\beq
 \hbM^2 |p;\T\ra =
 \left\{ \begin{array}{ll}
   0&\hbox{if } \t_{1}=\t_{2}=\t_{3};\\
   2|p;\T\ra & \hbox{otherwise.} \end{array}\right.
\label{3.9}
\eeq
Therefore each state $|p;\T\ra$ carries angular momentum $0$ or $1$. If
$p>2$, $W(p)$ decomposes as $(1)\oplus (3)\oplus(3)\oplus(1)$ with respect to
the $so(3)$ subalgebra of $sl(1|3)$; herein $(1)$ is a 1-dimensional subspace
with $M=0$ and $(3)$ is a 3-dimensional subspace with $M=1$. For $p=2$,
$W(2)=(1)\oplus (3)\oplus(3)$, and for $p=1$ one has $W(1)=(1)\oplus (3)$.

\section{On the position and momentum of the oscillating particle}
\setcounter{equation}{0}

The results in the previous section are not very precise about the
position of the oscillating particle in one of its stationary
states $\psi_q$ or $|p;\T\ra$: the only conclusion is that the
particle is localized on a sphere with radius $\varrho_q$.

We shall 
first investigate the probabilistic distribution of the particle
on the sphere corresponding to the states $|p;\T\ra$ or $\psi_q$.
In particular we shall show, with respect to measurements of
$\hat{R}_1$, $\hat{R}_2$ and $\hat{R}_3$, that  in the stationary
states $|p;\T\ra$ the particle can be found at only 8 points on
the sphere (we call them ``nests'') with radius $\varrho_q$,
see~(\ref{3.7}), and the number of such nests is even less in the
atypical cases $p=1$ and $p=2$.

The main tool to obtain these results is based
on the observation that the set of operators
\beq
{\hat H},~ {\hbR}^2,{\hbP}^2,~\hR_1^2,~\hR_2^2,~\hR_3^2,
~\hP_1^2,~\hP_2^2,~\hP_3^2,
\label{4.1}
\eeq
mutually commute and therefore  can be diagonalized
simultaneously. Observe that ${\hR}_k^2$ and ${\hP}_k^2$ and more
generally all even elements are independent of the time $t$,
which is why we do not write ${\hR}_k(t)^2$ and ${\hP}_k(t)^2$.

The sequence~(\ref{4.1}) contains in fact only 3
independent integral of motions, for instance $\hR_1^2$,
$\hR_2^2$ ,$\hR_3^2$, since
\beq
\hP_k^2 = m^2\omega^2 \hR_k^2,\ {\hat H}=m \omega^2{\hbR}^2,\
{\hbP}^2= m^2\omega^2 {\hbR}^2, \ \hbox{and }\
{\hbR}^2=\hR_1^2+\hR_2^2+\hR_3^2.
\label{4.2}
\eeq

All these are Hermitian operators in $W(p)$. Hence we can choose a
basis consisting of common eigenvectors to all of them.
In this case, we are lucky in the sense that all these operators
are already diagonal in the $\T$-basis.

At this point it is convenient to introduce dimensionless notation
for the energy, the coordinates and the momenta:
\beq
{\hat \e}={2\over \o\hbar}{\hat H}, \quad
{\hat r}_i(t)={\sqrt {2m\o\over \hbar}}\hR_i(t),\quad
{\hat p}_i(t)={\sqrt{2\over m\o \hbar}}\hP_i(t), \quad i=1,2,3.
\label{4.3}
\eeq
Then $\hr_i^2=\hp_i^2$, $i=1,2,3$ and
\beq
{\hat \e}={\hbr}^2={\hbp}^2~ =\sum_{i=1}^3\{A_i^+,A_i^-\}.
\label{4.4}
\eeq

\subsection{The basis vectors of $W(p)$ with $p>2$ (typical case)}

For $p>2$ all state spaces $W(p)$ of the system are $8$-dimensional.
The following holds:
\beq
 \hr_k^2 |p;\T\ra = \hp_k^2 |p;\T\ra =
  (p-q+\t_k) |p;\T\ra,~~k=1,2,3.
\label{4.5}
\eeq

What are the conclusions, which we can draw from
Eqs.~(\ref{4.5})? Let us answer this question first for
one particular state, e.g.\ $|p;1,1,0\ra$.
If measurements of the observables corresponding to
${\hbr}^2$, $\hr_1^2$, $\hr_2^2$, $\hr_3^2$
are performed, then according to
postulate {\bf P2} they will give the eigenvalues of these
operators, namely
\beq
{r}^2=3p-4, ~r_1^2=r_2^2=p-1, ~r_3^2=p-2.
\label{4.6}
\eeq
Moreover since the operators
${\hbr}^2$, $\hr_1^2$, $\hr_2^2$, $\hr_3^2$ commute
the results~(\ref{4.6}) can be measured simultaneously.
The latter means that if several measurements of the coordinates are
performed, then they will discover all of the time that
the particle is accommodated in one of 8 nests with coordinates
\beq
r_1=\pm \sqrt{p-1},~~~ r_2=\pm \sqrt{p-1},~~~ r_3=\pm\sqrt{p-2},
\label{4.7}
\eeq
of a sphere with radius $\rho=\sqrt{3p-4}$.

Similarly, the measurements of
the projections of the
momenta will give (due to~(\ref{4.5})):
\beq
p_1=\pm \sqrt{p-1},~~ p_2=\pm \sqrt{p-1},~~ p_3=\pm\sqrt{p-2}.
\label{4.8}
\eeq

The generalization of this result to any $\T$-state is evident:

\begin{conc}
If the system is in one of the $\T$-basis states
$|p;\T\ra$ then  measurements of
$r_1$, $r_2$ and $r_3$ imply that  the oscillating particle can be
found  in  no more than 8 nests with coordinates
\beq
 r_1=\pm \sqrt{p-q+\t_1},~~r_2= \pm\sqrt{p-q+\t_2},~~
r_3=\pm \sqrt{p-q+\t_3},
\label{4.9}
\eeq
on a sphere with radius $\rho_q=\sqrt{3p-2q}$. The measured
values of the momenta can take also only 8 different values,
\beq
 p_1=\pm \sqrt{p-q+\t_1},~~p_2= \pm\sqrt{p-q+\t_2},~~
p_3=\pm \sqrt{p-q+\t_3}.
\label{4.10}
\eeq
\end{conc}

Conclusion~2 significantly enhances the properties of the WQO
known so far, and collected in Conclusion~1. The particle is not
just anywhere on the sphere. In every $\T$-state $|p;\T\ra$ the
particle can be spotted in no more than 8 points of the sphere
with radius $\rho_q$. This is what we can say so far. What we
cannot say yet is whether some of these nests are not
forbidden for ``visits'' or what is the probability of finding the
particle in any one of them.

In order to investigate this last question we
shall need the eigenvectors and the eigenvalues of all the operators
of the coordinates and of the momenta. Before that a short remark
related to the properties of any WQS will be in order.

Let $\cal O$ be an observable and let  $x_1,\ldots, x_n$ be an
orthonormed basis of eigenvectors of $\hat O$: ${\hat O}x_i=O_i
x_i$. Assume that the system  is in a state $\psi=\a_1 x_1 +\ldots
+ \a_n x_{n}$ normalized to 1. Postulate {\bf P2} tells us that the
expectation value $\la\hat O\ra_\psi$ of the observable $\cal O$
in the state $\psi$ is
\beq
\la {\hat O}\ra_\psi=(\psi,{\hat O}\psi)= |\a_1|^2 O_1+\ldots
+ |\a_n|^2 O_n.
\label{4.11}
\eeq
It follows that $|\a_i|^2$ gives the probability of  measuring
the eigenvalue $O_i$ of the operator $\hat O$.
This is just {\em the superposition principle} of quantum mechanics. The conclusion
is that this principle holds for any WQS.

Thus in order to examine the probability for the particle to be
in one of the 8 nests, one has to introduce as a first step
an $\hr_k$-basis, namely an orthonormal basis of eigenvectors of
$\hr_k$  for any $k=1,2,3$. The second step is to express the
$\T$-basis via the $\hr_k$-basis for any $k=1,2,3$, and to apply
the superposition principle.

One has to proceed in a similar way in order to examine the
probability for the particle to have each one of the possible
values of momentum.

Let us define, for any $k\in\{1,2,3\}$ and any $\T$
satisfying~(\ref{2.21}), the following vectors in $W(p)$:
\beq
v_k(\T) = {1\over \sqrt{2}}\bigl( |p;\T_{\t_k=0}\ra +
(-1)^{\t_1+\cdots + \t_k} e^{-i\omega t} |p;\T_{\t_k=1}\ra\bigr).
\label{4.12}
\eeq
Herein, $\T_{\t_k=0}$ stands for the $\T$-value specified by the
left hand side of~(\ref{4.12}) in which $\t_k$ is replaced by~$0$ (and similarly
for $\T_{\t_k=1}$). Thus $v_k(\Theta)$ depends on $\theta_k$ only
through the sign factor $(-1)^{\t_1+\cdots + \t_k}$. A careful
computation shows that these (time-dependent) vectors $v_k(\T)$
constitute an orthonormal  basis of eigenvectors of $\hr_k(t)$ in
$W(p)$:
\beq
\hr_k(t)v_k(\T) = (-1)^{\t_k} \sqrt{p-q+\t_k} v_k(\T).
\label{4.13}
\eeq
The physical  interpretation of each eigenvector $v_k(\T)$
is clear (Postulate {\bf P2}): if (at the time $t$)
the oscillating particle is in a state $v_k(\T)$ then its
$k$-th coordinate is $(-1)^{\t_k} \sqrt{p-q+\t_k}$.

The inverse relations of~(\ref{4.12}) are also easy to write down:
\beq
|p;\T\ra = {1\over\sqrt{2}} (-1)^{(\t_1+\cdots+\t_{k-1})
\t_k} e^{i\omega t \t_k}
\bigl( v_k(\T_{\t_k=0}) + (-1)^{\t_k} v_k(\T_{\t_k=1})\bigr).
\label{4.14}
\eeq

The main observation needed is that in the inverse transformations
(\ref{4.14}) only two different vectors $v_k$ appear, each with a coefficient
of which the square modulus is 1/2.
In order to understand the importance of this observation,
consider an example, say $|p;1,1,0\ra$.
The expansion of this vector in the $\hr_k(t)$ eigenvectors
(for $k=1,2,3$) reads:
\begin{subeqnarray}
|p;1,1,0\ra &=& {e^{i\omega t}\over\sqrt{2}}\bigl(v_1(0,1,0)-v_1(1,1,0)\bigr)
 \slabel{4.15a}\\
& =& -{e^{i\omega t}\over\sqrt{2}}\bigl(v_2(1,0,0)-v_2(1,1,0)
\bigr) \slabel{4.15b}\\
& =& {1\over\sqrt{2}}\bigl(v_3(1,1,0)+v_3(1,1,1)\bigr).
 \slabel{4.15c}
\label{4.15}
\end{subeqnarray}
We see that the coefficients  of $v_1(0,1,0)$ and
$v_1(1,1,0)$ are equal up to a sign, and moreover their square
modulus is $1/2$. Therefore the superposition principle asserts
that with equal probability 1/2 the first coordinate $r_1$ of the
particle is either $+\sqrt{p-1}$ or  $-\sqrt{p-1}$. In other
words, the probability of finding the particle somewhere in the
four nests above the $r_2r_3$-plane is 1/2; and the probability to
find the particle somewhere in the four nests below the
$r_2r_3$-plane is also 1/2. Let us underline that this conclusion
is time independent. The time dependent basis which we have used
in order to derive it was playing only an intermediate role.

By means of the same arguments, using (\ref{4.15b}),
(\ref{4.15c}) and (\ref{4.13}),
one concludes that also with probability $1/2$ the second
coordinate $r_2$ and the third coordinate $r_3$ of the particle
take values $\pm \sqrt{p-1}$ and $\pm \sqrt{p-2}$, respectively
for the state $|p;1,1,0\ra$.

Taking the three results (about the probabilities for $r_1$,
$r_2$ and $r_3$) together does however not lead to a unique
solution for the probabilities to find the particle in a
particular nest.
Indeed, there are 8 probabilities to be
determined (one for each nest). From (\ref{4.15a}) we have deduced that
the sum of four of them (above the $r_2r_3$-plane) is 1/2, and the
sum of the remaining four (below the $r_2r_3$-plane) is also 1/2; so
this yields 2 linear relations for the 8 unknown probabilities.
Similarly, (\ref{4.15b}) and (\ref{4.15c}) each yield 2 linear relations. So
in total there are 6 linear relations in 8 unknowns. A more
detailed investigation even shows that only 4 of the 6 linear
relations are independent. This leads to the conclusion that the
probability of the particle being found in each nest cannot be
determined by the present considerations: there remain certain
degrees of freedom.

We have made this analysis for the example $|p;1,1,0\ra$, but
from (\ref{4.14}) it is clear that this conclusion generalizes to the
case of  all $\T$-states.
This follows from the fact that in the
inverse transformations (\ref{4.14}) only two different vectors $v_k$
appear, each with a coefficient of which the square modulus is
1/2.

We summarize the results in the next proposition.

\begin{prop}
If the system is
in one of the $\T$-basis states  $|p;\T\ra$, then measurements of
the position of the oscillating particle  are such that it can
only be observed to occupy one of  the 8 nests with
coordinates
\beq
 r_{k\pm}=\pm \sqrt{p-q+\t_k},~~k=1,2,3,
\label{4.16}
\eeq
on a sphere of dimensionless radius $\rho_q=\sqrt{3p-2q} $.
The probability ${\cal P}(\pm \pm \pm)$ of finding  the
particle in the nest with coordinates
$(r_{1\pm},r_{2\pm},r_{3\pm})$ cannot be determined. However, the
 probability of finding the particle somewhere in the four nests
with first coordinate equal to $r_{1+}$ is 1/2, and of  finding
it somewhere in the four nests with first coordinate equal to
$r_{1-}$ is also 1/2. The same holds for the second and third
coordinates.

The measurement of the momentum of the particle can take one of
the eight values
\beq
 p_{k\pm}=\pm \sqrt{p-q+\t_k},~~k=1,2,3.
\label{4.17}
\eeq
Again, the individual probabilities for each of the eight
possible values of the momenta cannot be determined; but the
the probability of having  a fixed component $p_{k+}$ is 1/2, and
that of a fixed component $p_{k-}$ is also 1/2 ($k=1,2,3$).
\end{prop}

The proof of the second part of Proposition~2, related to the
probabilities of the 8 possible values (\ref{4.17}) for the measurement
of the  momentum of the particle, is essentially the same as for
the coordinates. This time however, one has to use the orthonormal
basis of eigenvectors of $\hp_k(t)$, given by:
\beq
\tilde v_k(\T) = {1\over \sqrt{2}}\bigl( |p;\T_{\t_k=0}\ra - i
(-1)^{\t_1+\cdots + \t_k} e^{-i\omega t} |p;\T_{\t_k=1}\ra\bigr),
\label{4.18}
\eeq
with
\beq
\hp_k(t)\tilde v_k(\T) = (-1)^{\t_k} \sqrt{p-q+\t_k}
 \tilde v_k(\T).
\label{4.19}
\eeq
The inverse relations of (\ref{4.18}) are:
\beq
|p;\T\ra = { i^{\t_k}\over\sqrt{2}}
(-1)^{(\t_1+\cdots+\t_{k-1})\t_k} e^{i\omega t \t_k}
\bigl( \tilde v_k(\T_{\t_k=0}) + (-1)^{\t_k} \tilde v_k(\T_{\t_k=1})\bigr).
\label{4.20}
\eeq

The properties deduced in this subsection are summarized in
Figure~1.

\subsection{The basis vectors of $W(p)$ for $p\leq 2$ (atypical cases)}

So far we have considered the properties of almost all
state spaces $W(p)$. There are only two more cases left, namely
those with $p=1$ and $p=2$. We shall see in this section that
some of their physical properties are very different
from those of the typical cases, considered above.

\subsubsection{The state space $W(2)$}

For $p=2$, the state space $W(2)$ is 7-dimensional, since $|p;1,1,1\ra=0$.
Equations (\ref{4.5}) remain valid for all admissible $\T$-values (that is,
for all $\T$ with $\T\ne(1,1,1)$).
This implies that also Conclusion~2 (with equations (\ref{4.9}) and (\ref{4.10}))
remains valid for the admissible $\T$-values.
In this case, it is interesting to note that for the states with
$q=p=2$, one of the operators $\hr_k^2$ has zero eigenvalue.
For example, for the state $|2;1,1,0\ra$ one finds
\beq
r_1^2=r_2^2=1,\qquad r_3^2=0.
\label{4.28}
\eeq
Thus the third coordinate of the particle is zero. Also $p_3$,
the third component of the momentum, is zero. So the system
becomes flat, and the particle is ``polarized'' so as to lie in the
$r_1r_2$-plane. The oscillator behaves as a two-dimensional
object. The coordinates of the possible nests for this state are
$(r_1,r_2,r_3)= (\pm 1,\pm 1,0)$. So there are 4 nests where
the particle can be found (see Figure~2, where a complete picture
of the possibe $|p;\T\ra$ states is given); similarly, it can have only four
different momenta.

The conclusions about the probabilities, formulated in Proposition~2,
remain valid, but should be modified appropriately for the
lowest energy states with $q=2$.

\subsubsection{The state space $W(1)$}

The state space $W(1)$ is 4-dimensional.
The admissible $\T$-values have $\t_1+\t_2+\t_3\leq 1$.
For these admissible $\T$-values, (\ref{4.5}) and Conclusion~2
remain valid.
In this case, the interesting states are those with lowest
energy with $q=p=1$.
For these states, two of the operators $\hr_k^2$ have zero eigenvalue.
For example, for the state $|1;1,0,0\ra$ one finds
\beq
r_1^2=1,\qquad r_2^2=r_3^2=0.
\label{4.29}
\eeq
The coordinates of the two possible nests for this state are
$(r_1,r_2,r_3)=(\pm 1,0,0)$. Similarly, $p_2=p_3=0$ for this
state. So the oscillating system becomes one-dimensional, the
particle is ``polarized'' along the $r_1$-axis (see Figure~3,
where a copmplete picture of all $|p;\T\ra$ states is given).

For $q=1$ the considerations about probabilities of finding 
the particle in one of the two nests lead to a
unique solution. For each of the states $|1;\T\ra$ with
$q=p=1$, this probability is $1/2$. These nests are at the opposite poles on a
sphere with radius $1$.

\subsection{Arbitrary vectors of $W(p)$}

So far we were studying mainly the properties of the $\T$-states.
Here we proceed to consider some properties of the coordinates and momenta
for an arbitrary state $x\in W(p)$ and for any representation label $p$.

An arbitrary vector $x$ from the state space $W(p)$ can be
represented as
\beq
x=\sum_{\t_{123}\le p} \a (\t_1,\t_2,\t_3)|p;\t_1,\t_2,\t_3\ra,
\label{4.30}
\eeq
where $\a (\t_1,\t_2,\t_3)$ are any complex numbers, such that
\beq
\sum_{\t_{123}\le p} |\a (\t_1,\t_2,\t_3)|^2=1.
\label{4.31}
\eeq
Above and throughout
\beq
\t_{ijk}=\t_i +\t_j +\t_k
\label{4.32}
\eeq
and $\sum_{\t_{123}\le p}$ denotes a sum
over all $\t_1,\t_2,\t_3\in \{ 0,1\} $ with the additional
restriction $\t_1+\t_2+\t_3\le p$.
We shall be using also the polar form of  $\a (\t_1,\t_2,\t_3)$
\beq
\a(\T)=|\a(\T)|e^{i \varphi(\T)},
\label{4.33}
\eeq
where
$\a(\T)=\a(\t_1,\t_2,\t_3)$ and $\varphi(\T)=\varphi(\t_1,\t_2,\t_3)$.

The possible coordinates  (and momenta) of the oscillator, in an
arbitrary state $x$, follows from the previous discussions and
the superposition principle. For clarity, let us formulate it for
$W(p)$ with $p>2$. Then a measurement of the position of the
particle in the state $x$ will yield one of the 64 possible nests
(see Figure~1)
$$
(r_1,r_2,r_3)= (\pm \sqrt{p-q+\t_1}, \pm\sqrt{p-q+\t_2}, \pm
\sqrt{p-q+\t_3}), ~~ {\rm with }~~ q=\t_1+\t_2+\t_3.
$$

The probability of finding  the particle somewhere in the eight
nests associated with $|p;\t_1,\t_2,\t_3\ra$ is given by
$|\alpha(\t_1,\t_2,\t_3)|^2$, but the probability for each nest
separately cannot be determined. Similarly, an arbitrary state
$x$ of $W(2)$ can be in 44 possible nests (see Figure~2); an arbitrary state
$x$ of $W(1)$ can be found in 14 possible nests (see Figure~3).

In order to give more properties of the position probabilities,
it is again necessary to expand the general state $x$ in terms of
the orthonormalized eigenstates of $\hr_k(t)$. Let us do it
here explicitly for $p>2$; (\ref{4.30}) and (\ref{4.14}) imply:
\beq
x = \sum_\T {1\over\sqrt{2}} \Bigl(
\a(\T_{\t_k=0}) + (-1)^{\t_1+\cdots+\t_{k-1}} e^{i\omega t}
\a(\T_{\t_k=1}) \Bigr) v_k(\T).
\label{4.34}
\eeq
Then the square modulus of the coefficient in
front of $v_k(\T)$ yields the probability of the
particle in the state $x$
being observed to
have $\r_k(t)$ eigenvalue $(-1)^{\t_k}\sqrt{p-q+\t_k}$.

Many of our formulas to be presented later will look rather
complicated in the general state $x$, so sometimes we shall
concentrate on a particular example of such a normalized state
which carries all the main features of the general picture. We
take as our standard example one of the simplest non-stationary
states (we assume that $p>2$, but most of the results, apart from
the number of the nests hold for $p=1$ and $2$)
\beq
z= {1\over{\sqrt 2}} |p;0,0,0\ra + {1\over{\sqrt 2}}|p;0,0,1\ra.
\label{4.35}
\eeq

The consideration of such an example will help to understand
some of the peculiar features of the WQO in a general state.
Let us explicitly deduce what can be said about the position of
the particle when the system is in the state $z$.
First of all, only two $\T$-states are involved in (\ref{4.35}),
each of these $\T$-states corresponding to 8 nests.
All of these nests are different,
so the particle can be detected in 16 possible nests.
The probability of detecting the particle somewhere in the 8 nests
corresponding to $|p;0,0,0\ra$ or to  $|p;0,0,1\ra$)
is $1/2$; these probabilities are just the square
moduli of the coefficients in (\ref{4.35}).

The state (\ref{4.35}) of the oscillator corresponds to a configuration
in which 8 nests have value of $r_3=\sqrt{p}$ and
the other 8 states have $r_3=-\sqrt{p}$ (see Figure~1). We cannot determine the
probabilities of the 16 nests separately, but we can draw
conclusions about the probability ${\cal P}(r_3=\pm \sqrt{p})$ of
detecting the particle in the nests with a given $r_3=\pm
{\sqrt{p}}$. To this end consider the expansion of the state $z$
in terms of the eigenvectors of $\hr_3(t)$:
\beq
z={1\over 2}(1+e^{i\omega t})v_3(0,0,0)+
{1\over 2}(1-e^{i\omega t})v_3(0,0,1).
\label{4.36}
\eeq
Then the square moduli of the coefficients give the probabilities
of finding the particle in the nests with a particular
$r_3$-value. So we find:
\begin{subeqnarray}
&& r_3=-\sqrt{p} \quad  {\rm with~probability}~~
  {\cal P}(r_3=- \sqrt{p})= {1-\cos(\omega t)\over 2}, \slabel{4.37a}\\
&& r_3=~~\sqrt{p} \quad {\rm with~probability}~~
 {\cal P}(r_3=\sqrt{p})= {1+\cos(\omega t)\over 2}. \slabel{4.37b}
\label{4.37}
\end{subeqnarray}

Equations (\ref{4.37}) describe an interesting new phenomenon, which
does not show up whenever the oscillator is in one of the
$\T$-basis states or more generally in any stationary state (\ref{3.2}).
As it should be $ {\cal P}(r_3=- \sqrt{p})+ {\cal
P}(r_3=\sqrt{p})=1$. But the probabilities are time dependent.
There is an oscillation of the probabilities: the probabilities
for  the particle to be found in the nests  either with
$r_3=\sqrt{p}$ or with $r_3=-\sqrt{p}$ vary from zero to one.

Contrary to this the probabilities of  finding  the particle in
the nests with a fixed $r_1$-value, or with a fixed $r_2$-value
are time independent. This follows from the expansion of the
$z$-state in terms of the eigenvectors of $\hr_1(t)$ and
$\hr_2(t)$, which yields:
\beq
{\cal P}(r_1=\sqrt{p})={\cal P}(r_1=-\sqrt{p})={\cal P}(r_1=
\sqrt{p-1})={\cal P}(r_1=-\sqrt{p-1})={1\over 4},
\label{4.38}
\eeq
and
\beq
{\cal P}(r_2= \sqrt{p})={\cal P}(r_2=-\sqrt{p})={\cal P}(r_2=
\sqrt{p-1})={\cal P}(r_2=-\sqrt{p-1})={1\over 4}.
\label{4.39}
\eeq

In the case of $p=1$ (\ref{4.36}) and (\ref{4.37}) still  hold, so the
oscillations of the probabilities along the $r_3$-axis  remain
unaltered. In this case however two of the 10 nests, those
associated with $|1;0,0,1\ra $, are on the third axis, which
yields:
\beq
{\cal P}(r_1= 1)={\cal P}(r_1= -1)={1\over 4}, \quad {\cal
P}(r_1=0)= {1\over 2},
\label{4.40}
\eeq
and
\beq
{\cal P}(r_2= 1)={\cal P}(r_2= -1)={1\over 4}, \quad {\cal
P}(r_2=0)= {1\over 2}.
\label{4.41}
\eeq
Based on (\ref{4.37})-(\ref{4.39}) we can compute the average values of the
coordinates in the state $z$:
\begin{subeqnarray}
&& \la \hat{r}_3(t)\ra_z =  -\sqrt{p}\  {1-\cos(\omega t)\over 2} +
\sqrt{p}\ {1+\cos(\omega t)\over 2}=
\sqrt{p}\,\cos(\omega t).     \slabel{4.42a}\\
&& \la \hat{r}_1(t)\ra_z = \la \hat{r}_2(t)\ra_z = 0. \slabel{4.42b}
\label{4.42}
\end{subeqnarray}

\section{Mean trajectories and standard deviations of
positions and momenta}
\setcounter{equation}{0}

Now that we have discussed some properties of the position
operators in more general states, let us next compute the mean
trajectories or time dependent expectation values of the
coordinates and momenta and their standard deviations in a
general state $x$. We shall then specify our results to the
stationary states $\psi_q$. We shall indicate also (Conclusion~3)
that for the WQS there exists no (nontrivial) Heisenberg
uncertainty relations.

For the mean trajectory of the coordinates in an arbitrary
state $x$ we obtain:
\bea
\la\hr_{k}(t)\ra_x &=& (x,\hr_k(t) x)=\sum_{\t_{123}\le p}
(-1)^{\t_1+\cdots + \t_{k-1}} \sqrt{p-q+\t_k}
|\alpha(\T_{\t_k=0}) \alpha(\T_{\t_k=1})| \nn\\
&& \times \cos\bigl(\omega t
-\varphi(\T_{\t_k=0})+\varphi(\T_{\t_k=1})\bigr),
\label{5.1}
\eea
where as in (\ref{4.12}) $\T_{\t_k=0}$ stands for the $\T$-value in
which $\t_k$ is replaced by~$0$ (and similarly for $\T_{\t_k=1}$).
Observe that in (\ref{5.1}), the contributions come in equal pairs;
e.g.\ for $k=1$, the contribution coming from $\T=(0,\t_2,\t_3)$
is the same as that coming from $\T=(1,\t_2,\t_3)$, since
$\sqrt{p-q+\t_k}$ is independent of $\t_k$. Similarly, one finds:
\bea
\la\hp_{k}(t)\ra_x &=& (x,\hp_k(t) x)= - \sum_{\t_{123}\le
p} (-1)^{\t_1+\cdots + \t_{k-1}} \sqrt{p-q+\t_k}
|\alpha(\T_{\t_k=0}) \alpha(\T_{\t_k=1})| \nn\\
&& \times \sin\bigl(\omega t
-\varphi(\T_{\t_k=0})+\varphi(\T_{\t_k=1})\bigr).
\label{5.2}
\eea
For instance, for our standard example (\ref{4.35}), we find
\begin{subeqnarray}
&&\la \hr_1(t)\ra_z = 0, \quad \la \hr_2(t)\ra_z = 0, \quad
 \la \hr_3(t)\ra_z = {\sqrt{p}} \cos(\omega t), \slabel{5.3a}\\
&&\la \hp_1(t)\ra_z =0, \quad  \
 \la \hp_2(t)\ra_z = 0, \quad
 \la \hp_3(t)\ra_z = -{\sqrt{p}} \sin(\omega t). \slabel{5.3b}
\label{5.3}
\end{subeqnarray}

Note that each term in the right hand sides of
(\ref{5.1})-(\ref{5.2}) contains  multiples
\beq
\a (\t_1,\t_2,\t_3)\a ({\tilde \t}_1,{\tilde \t}_2,{\tilde \t}_3)
\quad {\rm such~that}\quad \t_1+\t_2+\t_3\ne
{\tilde \t}_1+{\tilde \t}_2+{\tilde \t}_3.
\label{5.4}
\eeq
Therefore the right hand sides of  (\ref{5.1})-(\ref{5.2}) vanish if the system is in a
stationary state $\psi_q$. The latter stems from the observation
that in the stationary states, see (\ref{3.2}), $x$ is a linear
combination of basis states $|p;\t_1,\t_2,\t_3\ra$ with fixed
$\t_1+\t_2+\t_3$, namely all non-zero coefficients $\a
(\t_1,\t_2,\t_3)$ in (\ref{4.30}) have one and the same
$q=\t_1+\t_2+\t_3$. Thus we have

\begin{conc}
The mean trajectories of the
coordinates and momenta vanish if the system is in a
stationary state $\psi_q$.
\end{conc}

In order to draw conclusions about the uncertainty of the
coordinates and momenta, more precisely about their standard deviations,
we also need the mean square deviations of
$\hr_k$ and $\hp_k$, $k=1,2,3$.
It follows from (\ref{4.5}) that
\beq
\la \hr_k(t)^2 \ra_x = \la \hp_k(t)^2 \ra_x = \sum_{\t_{123}\le p}
 (p-q+\t_k) |\alpha(\T)|^2.
\label{5.5}
\eeq

Recall that the general definition of the standard deviation
$\Delta X$ of an observable $X$ in a state $x$ is given by
\beq
\Delta X_x =   \sqrt{\la X^2\ra_x - \la X\ra_x^2}.
\label{5.6}
\eeq
So from the previous formulas one can write down the standard
deviation of $\hr_k(t)$ and $\hp_k(t)$ in an arbitrary state $x$:
\begin{subeqnarray}
\Delta \hr_k(t)_x  &=&  \Big[\sum_{\t_{123}\le p}
 (p-q+\t_k) |\alpha(\T)|^2 -
 \Bigl(\sum_{\t_{123}\le p}
 (-1)^{\t_1+\cdots+\t_{k-1}} \sqrt{p-q+\t_k} \nn\\
&& \times |\alpha(\T_{\t_k=0}) \alpha(\T_{\t_k=1})|
\cos\bigl(\omega t
-\varphi(\T_{\t_k=0})+\varphi(\T_{\t_k=1})\bigr)
\Bigr)^2\Big]^{1/2}, \slabel{5.7a} \\
\Delta \hp_k(t)_x &=& \Big[ \sum_{\t_{123}\le p}
 (p-q+\t_k) |\alpha(\T)|^2 - \Bigl(\sum_{\t_{123}\le p}
 (-1)^{\t_1+\cdots+\t_{k-1}} \sqrt{p-q+\t_k} \nn\\
&& \times |\alpha(\T_{\t_k=0}) \alpha(\T_{\t_k=1})|
\sin\bigl(\omega t -\varphi(\T_{\t_k=0})+\varphi(\T_{\t_k=1})\bigr)
\Bigr)^2\Big]^{1/2}. \slabel{5.7b}
\label{5.7}
\end{subeqnarray}
Because of the double products in the expansion of the square,
(\ref{5.7}) cannot be simplified further for an arbitrary state vector
$x$.

Let us just observe that in any one of the stationary states
$\psi_q$ the standard deviations become very simple and are time independent:
\begin{subeqnarray}
&& \Delta \hr_j(t)_{\psi_0} =\Delta \hp_j(t)_{\psi_0} = \sqrt{p}, \slabel{5.8a}\\
&& \Delta \hr_j(t)_{\psi_1} = \Delta \hp_j(t)_{\psi_1}
=  \sqrt{p-1+|\alpha(\t_j=1,\t_k=\t_l=0)|^2}\ge \sqrt{p-1}, \slabel{5.8b}\\
&& \Delta \hr_j(t)_{\psi_2} =\Delta \hp_j(t)_{\psi_2} =
 \sqrt{p-1-|\alpha(\t_j=0,\t_k=\t_l=1)|^2}\ge \sqrt{p-2}, \slabel{5.8c}\\
&& \Delta \hr_j(t)_{\psi_3} = \Delta \hp_j(t)_{\psi_3} =
\sqrt{p-2}, \slabel{5.8d}
\label{5.8}
\end{subeqnarray}
where $j\ne k \ne l \ne j \in\{1,2,3\}$.

Although formulas (\ref{5.7}) look complicated, they are
easy to apply. For instance, for our standard example (\ref{4.35}),
we find
\bea
&& \Delta \hr_1(t)_z =\Delta \hp_1(t)_z = \Delta
\hr_2(t)_z =\Delta \hp_2(t)_z = \sqrt{{2p-1 \over 2}}, \nn\\
&& \Delta \hr_3(t)_z = {\sqrt p}|\sin(\o t)|,\quad \Delta \hp_3(t)_z =
{\sqrt p}|\cos(\o t)|. \label{5.9}
\eea

Equations (\ref{5.7}) can be used for independent verification of some
of the properties  of the WQOs. Consider for instance the state
$|2;1,1,0\ra$. We know, see (\ref{4.28}), that in this state the
particle is polarized in the $r_1r_2$-plane, both $r_3=0$ and
$p_3=0$. Equation (5.7) confirms this:
\beq
\Delta \hr_3(t)_y=\Delta \hp_3(t)_y=0 \quad {\rm in~the~state}
~~~~y=|2;1,1,0\ra.
\label{5.10}
\eeq

Let us note more generally that for any $p$ and $k=1,2,3$ there exists a state
$x_k$ and a time $t$ such that $\Delta \hr_k(t)_{x_k}=0$
or $\Delta \hp_k(t)_{x_k}=0$. For instance
\begin{subeqnarray}
&& \Delta \hr_1(0)_{x_1}=0 \quad {\rm for} \quad
x_1={1\over \sqrt 2}(|p;0,1,0\ra + |p;1,1,0\ra), \slabel{5.11a}\\
&& \Delta \hr_2(0)_{x_2}=0 \quad {\rm for} \quad
x_2={1\over \sqrt 2}(|p;0,0,1\ra + |p;0,1,1\ra), \slabel{5.11b}\\
&& \Delta \hr_3(0)_{x_3}=0 \quad {\rm for} \quad
x_3={1\over \sqrt 2}(|p;1,0,0\ra + |p;1,0,1\ra). \slabel{5.11c}
\label{5.11}
\end{subeqnarray}
As an immediate consequence we have

\begin{conc}
The position and momentum operators of a
WQO do not satisfy an uncertainty-like relation of the form
\beq
 \Delta \r_k(t)_x \Delta \p_k(t)_x \geq \gamma
\label{5.12}
\eeq
for any $\gamma>0$ holding simultaneously for all states $x$ of
the system at all times $t$ (as is the case for a CQO and, more
generally, for any canonical quantum system).
\end{conc}

If however, $x$ is any stationary state $\psi_q$, then in the
typical case with $p>2$, (\ref{5.8}) yields
\beq
 \Delta \r_k(t)_{\psi_q} = \Delta \p_k(t)_{\psi_q} \geq \sqrt{p-2}
\label{5.13}
\eeq
for all $k=1,2,3$. Therefore for any stationary state and any time
\beq
  \Delta \r_i(t)_{\psi_q} \Delta \r_j(t)_{\psi_q} \geq p-2,
  ~~~~
  \Delta \r_i(t)_{\psi_q} \Delta \p_j(t)_{\psi_q} \geq p-2,
  ~~~~
  \Delta \p_i(t)_{\psi_q} \Delta \p_j(t)_{\psi_q} \geq p-2,
\label{5.14}
\eeq
with $p-2>0$.

Returning to the case of an arbitrary state $x\in W(p)$,
uncertainty-like relations of the type (\ref{5.12}) certainly will
exist, but the uncertainty parameter $\gamma$ may be zero. They
can be derived from the general uncertainty relation~\cite{42}
\beq
     \Delta \F(t)_x \Delta \G(t)_x \geq {1\over2}
     \left|~\la~[\F(t),\G(t)]~\ra_x~\right|,
\label{5.15}
\eeq
that applies to any two Hermitian operators $\F$ and $\G$ for any
$x\in W(p)$.

Applying this in the case $\F(t)=\r_k(t)$ and $\G(t)=\p_k(t)$ and
the arbitrary state $x$ as defined in (\ref{4.30}) gives
\beq
   \Delta \r_k(t)_{x} \Delta \p_k(t)_{x} \geq
   \left|~ \sum_{\t_{123}\leq p} (-1)^{\theta_k} (p-q+\theta_k)
   |\alpha(\Theta)|^2
   ~\right|.
\label{5.16}
\eeq
It should be noted that the sign factors $(-1)^{\theta_k}$ are
such that cancellations may occur and may yield zero on the right
hand side, as is the case, for example, if $k=3$ and $x$ is the
state $x_3={1\over \sqrt 2}(|p;1,0,0\ra + |p;1,0,1\ra)$ introduced
in (\ref{5.11c}).

On the other hand  if (\ref{5.16}) is restricted to the stationary state
$\psi_0$, for example, then it yields a relation very similar,
when properly dimensionalized, to the Heisenberg uncertainty
relation, namely:
\beq
   \Delta \R_k(t)_{\psi_0} \Delta \P_k(t)_{\psi_0} \geq
   {{\hbar p}\over{2}}.
\label{5.17}
\eeq

Formulas of the type (\ref{5.16}) in the case of $\Delta \r_k(t)_{x}
\Delta \r_l(t)_{x}$, $\Delta \r_k(t)_{x} \Delta \p_l(t)_{x}$ and
$\Delta \p_k(t)_{x} \Delta \p_l(t)_{x}$ with $k\neq l$ are
however much more involved and will not be analyzed here systematically.
We just give a typical example illustrating the impact of the
non-commutative geometry:
\bea
\Delta \r_1(t)_{x} \Delta \r_2(t)_{x} &\geq & 
 {1\over 2} \left| \la~[\r_1(t),\r_2(t)]~\ra_x \right| \nn\\
&=&
\big| 2\sqrt{p(p-1)} |\alpha(0,0,0)\alpha(1,1,0)|
 \sin(2\omega t-\phi(0,0,0)+\phi(1,1,0))\nn\\
&& +2\sqrt{(p-1)(p-2)} |\alpha(0,0,1)\alpha(1,1,1)|
 \sin(2\omega t-\phi(0,0,1)+\phi(1,1,1))\nn\\
&& +(2p-1)|\alpha(1,0,0)\alpha(0,1,0)|
 \sin(\phi(1,0,0)-\phi(0,1,0))\nn\\
&& +(2p-3)|\alpha(1,0,1)\alpha(0,1,1)|
 \sin(\phi(1,0,1)-\phi(0,1,1))
   ~\big|~. \label{5.18}
\eea

\section{Comparison with the canonical quantum oscillator}
\setcounter{equation}{0}

{}From the discussions so far it becomes clear that the WQOs are
very different from the Bose canonical quantum oscillators (CQOs).
Therefore it is somewhat of a surprise that one can  establish a
one-to-one correspondence between some mean trajectories of the
3D CQO  and the $p=1$ mean trajectories of the WQO. This will be the
main topic of the present section.

In dimensionless units, see (\ref{4.3}), the coordinates $\br_k$
and momenta $\bp_k,~k=1,2,3$ of a 3D canonical oscillator
(\ref{2.5}) read:
\beq
\br_k(t)=b_k^+e^{~i \omega t}
+ b_k^-~e^{-i \omega t} ,
\quad \bp_k(t)=i~
(b_k^+e^{~i \omega t}
-~ b_k^-~e^{-i \omega t} ).
\label{6.1}
\eeq

Let us consider first a simple example. As a Bose analogue of our
standard state $z$ we set
\beq
{\bar z}= {1\over{\sqrt 2}} |0,0,0) + {1\over{\sqrt 2}}|0,0,1).
\label{6.2}
\eeq
It is a simple computation to show that the mean trajectories  of
the coordinates and momenta in the state ${\bar z}$ of the Bose
oscillator read:
\begin{subeqnarray}
&& \la \br_1(t)\ra_z = 0, \quad \la \br_2(t)\ra_z = 0, \quad
 \la \br_3(t)\ra_z =  \cos(\omega t), \slabel{6.3a}\\
&& \la \bp_1(t)\ra_z =0, \quad  \
 \la \bp_2(t)\ra_z = 0, \quad
 \la \bp_3(t)\ra_z = - \sin(\omega t). \slabel{6.3b}
\label{6.3}
\end{subeqnarray}
The above trajectories are the same as those of the WQO given in
(\ref{5.3}) provided in the latter that $p=1$. This was the first
indication that some of the trajectories of the WQO are the same
as those of the canonical Bose oscillator. The question is  how
far does this similarity go. In the next proposition we summarize
the results which we are able to establish.

\begin{prop}
To each $p=1$ mean trajectory in the
phase space of the Wigner quantum oscillator there corresponds an
identical trajectory of the 3D Bose canonical quantum oscillator.
\end{prop}

Let $p=1$. By a straightforward computation one shows
that the mean trajectory of the WQO in the state
\beq
x=\a(0,0,0)|1;0,0,0\ra+\a(1,0,0)|1;1,0,0\ra
+\a(0,1,0)|1;0,1,0\ra +\a(0,0,1)|1;0,0,1\ra
\label{6.4}
\eeq
is the same as the mean trajectory of the Bose oscillator
in the state
\beq
x^*=\a(0,0,0)^*|0,0,0)+\a(1,0,0)^*|1,0,0)+\a(0,1,0)^*|0,1,0)
+\a(0,0,1)^*|0,0,1).
\label{6.5}
\eeq
The $*$ in the right hand side of (\ref{6.5}) denotes complex conjugation.

Explicitly the mean trajectories corresponding to (\ref{6.4}) and
(\ref{6.5}) read:
\begin{subeqnarray}
&& \la\hr_{k}(t)\ra_x  = \la\br_{k}(t)\ra_{x^*}
 = 2  \ |\a (0,0,0)\a(0,0,0)_{\t_k=1}|\
  \cos\big(\omega t + \varphi(0,0,0)_{\t_k=1} -
  \varphi(0,0,0)\big), \nn\\
&& \slabel{6.6a}\\
&& \la\hp_{k}(t)\ra_x  = \la\bp_{k}(t)\ra_{x^*}
 = - 2  \ |\a (0,0,0)\a(0,0,0)_{\t_k=1}|\
  \sin\big(\omega t + \varphi(0,0,0)_{\t_k=1} -
  \varphi(0,0,0)\big), \nn\\
&& \slabel{6.6b}
\label{6.6}
\end{subeqnarray}
where $\a (0,0,0)_{\t_k=1}$ denotes $\a (1,0,0)$, $\a (0,1,0)$,
$\a (0,0,1)$ according as $k=1,2,3$, respectively.

However, although the standard deviations of the coordinates and
momenta of the WQO in a state $x$ and of the CQO in a state $x^*$
also look somewhat similar, they  are in fact different:
\bea
&{\rm WQO:}& \Delta \hr_k(t)_x =\Big[|\a(0,0,0)|^2 +
|\a(0,0,0)_{\t_k=1}|^2 \label{6.7} \\
&& -4 |\a(0,0,0)\a(0,0,0)_{\t_k=1}|^2
\cos^2(\omega t -\varphi(0,0,0) + \varphi(0,0,0)_{\t_k=1})
\Big]^{1/2}, ~~k=1,2,3.\nn\\
&{\rm CQO:}& \Delta \br_k(t)_{x^*} =\Big[1 +
2|\a(0,0,0)_{\t_k=1}|^2 \label{6.8} \\
&& -4 |\a(0,0,0)\a(0,0,0)_{\t_k=1}|^2
\cos^2(\omega t -\varphi(0,0,0) + \varphi(0,0,0)_{\t_k=1})
\Big]^{1/2}, ~~k=1,2,3. \nn
\eea
Our standard state $z$ provides   a good illustration of  the
difference:
\bea
&{\rm WQO:}& \Delta \hr_1(t)_z=\Delta \hr_2(t)_z= {1\over{\sqrt
2}},\quad \Delta \hr_3(t)_z=|\sin(\o t)|, \label{6.9}\\
&{\rm CQO:}& \Delta \br_1(t)_{z^*}=\Delta \br_2(t)_{z^*}= 1, \quad
\Delta \br_3(t)_{z^*}=\big(1+\sin^2(\o t)\big)^{1/2}. \label{6.10}
\eea
Let us go further and compare the standard deviations
corresponding to the state $y=|1;1,0,0\ra$ and its Bose
``partner'' $y^*=|1,0,0)$.
\bea
& {\rm WQO:}& \Delta \hr_1(t)_y=\Delta
\hp_1(t)_y=1,\quad \Delta \hr_k(t)_y=\Delta \hp_k(t)_y=0,~~k=1,2, \label{6.11}\\
& {\rm CQO:}& \Delta \br_1(t)_{y^*}=\Delta \bp_1(t)_{y^*}
=\sqrt{3},\quad \Delta \br_k(t)_{y^*}=\Delta \bp_k(t)_{y^*}=1,~~k=1,2.
\label{6.12}
\eea
Equation (\ref{6.11}) confirms that the oscillator in the state
$|1;1,0,0\ra$ ``lives'' on the $r_1$-axis. This is not the case
for the CQO,  either in the state $|1,0,0)$ or in any other
state, since the right hand side of  (\ref{6.8}) never vanishes, as is implied
also by the Heisenberg uncertainty relations.

We have compared also the sizes of the WQO and
CQO corresponding
to the basis states. For the CQO
\beq
{\br}(t)^2 = \sum_{k=1}^3
\Big((b_k^+)^2 e^{2i\omega t} +
\{b_k^+,b_k^-\} + (b_k^-)^2 e^{-2i\omega t}\Big)
\label{6.13}
\eeq
is not an integral of motion. However for  the average value of
$\br^2$ in the state $x^*$ we find
\beq
\la{\br}(t)^2\ra_{x^*}=5 - 2|\a(0,0,0)|^2.
\label{6.14}
\eeq
Thus we have
\bea
& {\rm WQO:}& \la{\hr}(t)^2\ra_{|1;0,0,0\ra}=3,~~
\la{\hr}(t)^2\ra_{|1;1,0,0\ra}=\la{\hr}(t)^2\ra_{|1;0,1,0\ra}=
\la{\hr}(t)^2\ra_{|1;0,0,1\ra}=1,  \label{6.15} \\
& {\rm CQO:}& \la{\br}(t)^2\ra_{|0,0,0)}=3,~~ \la{\br}(t)^2\ra_{|1,0,0)}=
\la{\br}(t)^2\ra_{|0,1,0)}= \la{\br}(t)^2\ra_{|0,0,1)}=5.
\label{6.16}
\eea
Thus only the states $|1;0,0,0\ra$ of the WQO and $|0,0,0)$ of the
CQO have one and the same space dimensions. This is perhaps not
surprising since only these states have  one and the same energy
${\epsilon}=3$ (in units of $\o \hbar/2$). The energy of the other
WQO states is 1, whereas for the other CQO states it is 5.

\section{Concluding remarks}
\setcounter{equation}{0}

It is clear that while alternative, non-canonical solutions to the
compatibility equations (\ref{2.4}) between Hamilton's equations and the
Heisenberg equations exist in our $sl(1|3)$ WQO model, they are
in several very important respects quite different from the
canonical solutions.

Firstly, each state space $W(p)$ of our one particle 3D WQO is
finite-dimensional; 8-dimensional in the case of typical
representations of $sl(1|3)$, and either $7$ or $4$-dimensional
in the case of atypical representations.

Secondly, both the energy and the angular momentum are quantized,
with equally spaced energy levels, all positive and with
separation ${1\over2}\hbar \omega$, and with the angular momentum
restricted to be $0$ or $1$. Since there are only a finite number
of energy levels, the energy is bounded. The degeneracies are
always either $3$ or $1$. The lowest energy state is
non-degenerate in all the typical cases, but degenerate in each of
the atypical cases.

Thirdly, the spectrum of coordinates is also quantized, to the
extent that in any stationary state measurements of the
coordinates $r_1$, $r_2$ and $r_3$ give values consistent only
with the particle being found at a finite number of possible
sites, namely the various nests that we have identified. In the
typical, $p>2$ case, the number of possible nests is $64$, while
in the atypical cases it is $44$ if $p=2$ and only $14$ if $p=1$.
In all cases the distance of the particle from the origin is
bounded and may take on only the values
$\sqrt{(\hbar/2m\omega)(3p-2q)}$ with $q\in\{0,1,2,3\}$ such that
$q\leq p$.

Fourthly, not only is the mean trajectory of the particle in any
stationary state zero, but there exist both typical and atypical
states for which the standard deviation of some coordinate $r_k$,
or some component of linear momentum $p_k$, is also zero. This
implies that for the WQO there exists no uncertainty relation
involving a non-zero uncertainty parameter $\gamma$ that applies
to all states at all times.

Fifthly, the atypical case is distinguished from the typical case
in possessing stationary states of dimension lower than three,
namely two-dimensional in the case $p=2$ and one-dimensional in
the case $p=1$.

Many of these non-standard results are a consequence of the fact
that the underlying geometry of this WQO model is
non-commutative. This means that their interpretation must be
undertaken carefully. In particular it should be stressed that it
is not possible to specify precisely the position of the particle.
Fortunately, the square operators $\hr_1^2$, $\hr_2^2$
and $\hr_3^2$ not only mutually commute but also commute with the
Hamiltonian. Their common eigenstates are the stationary states
$|p;\T\ra$, for which the eigenvalues of $\hr_1^2$, $\hr_2^2$
and $\hr_3^2$ are simultaneously  fixed to be either $p$, $p-1$ or $p-2$.
Thus the spectrum of the measured values of each of the
coordinates themselves, $r_1$, $r_2$ and $r_3$, is necessarily
restricted to the set of values $\pm\sqrt{p}$, $\pm\sqrt{p-1}$
and $\pm\sqrt{p-2}$. Any measurement of a coordinate, $r_1$ say,
results in one or other of the allowed positive or negative
values of $r_1$ with, as we have shown, equal probability,
leaving the signs of the other coordinates undetermined. Thus the
particle in a stationary state $|p;\T\ra$ has a certain
probability of being within one or other of the relevant nests,
but it is not to be thought of as localized in any particular
nest.

In this article we have restricted ourselves to the relatively
simple case of an $n=1$, single particle 3-dimensional WQO with a
relatively low, $2^3$-dimensional Fock space $W(p)$ associated
with typical irreducible representations of $sl(1|3)$, and even
lower dimensions for atypical representations. A very natural
next step is to generalize the results to the case of an $n$-body
3-dimensional WQO as introduced in~\cite{2}. In such a case the
dimension of the Fock space is $2^{3n}$-dimensional for typical
representations, and again lower dimensions for atypical
representations. The relevant calculations are somewhat intricate
and will be the subject of a separate article, in which the
restriction from the Lie superalgebra $sl(1|3n)$ to the simple Lie
algebra $so(3)$ of the rotation group plays a key role in the
determination of the possible angular momentum states of our
multiparticle system. It suffices to say at this stage that not
only are the energy and angular momentum quantized and bounded
but, in the corresponding stationary states of fixed energy and
angular momentum, so are the single particle coordinates and
components of linear momentum. As is to be expected the
corresponding nest structure is more complicated and we have to
contend with the relevant class $A$ statistics, and account for
the numbers of particles whose coordinates, when measured, can
coincide with those of the nests, as well as more complicated
patterns of degeneracy.

Further generalizations also come to mind. In particular the
class of irreducible representations considered here are those
specified by the parameter $p$. There exist other
finite-dimensional irreducible representations of $sl(1|3)$, and
more generally of $sl(1|3n)$, that are specified not just by a
single positive integer $p$, but by a partition or equivalently a
sequence of Kac-Dynkin indices~\cite{21}. These can be expected to
provide other interesting models of the Wigner quantum oscillator
in the one-particle case and, more particularly, in
multi-particle cases. At the same time it would be interesting to
explore in the same way other non-oscillator Wigner quantum systems.
For examples of this kind see~\cite{7,10,12}.

\section*{Acknowledgements}

TDP is  thankful to Prof.\ Randjbar-Daemi for the kind invitation
to visit the High Energy Section of the Abdus Salam International
Centre for Theoretical Physics and to Prof.~D.~Trifonov for the numerous
discussions.  NIS has been supported by a Marie Curie Individual Fellowship
of  the European Community Programme ``Improving the Human Research Potential
and the Socio-Economic Knowledge Base" under contract number
HPMF-CT-2002-01571. This work was supported also
by the Royal Society Joint Project Grant UK-Bulgaria H01R381 and
by NATO (Collaborative Linkage Grant).

\newpage

\begin{center}
\begin{tabular}{ccc}
 & $|p;0,0,0\rangle$ & \\[3mm]
 & \includegraphics{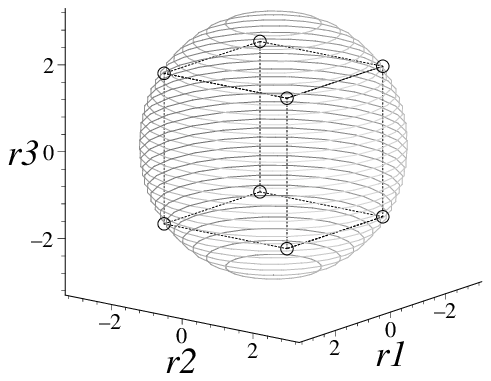} &  \\
$|p;1,0,0\rangle $ & $|p;0,1,0\rangle $ & $|p;0,0,1\rangle $\\[3mm]
\includegraphics{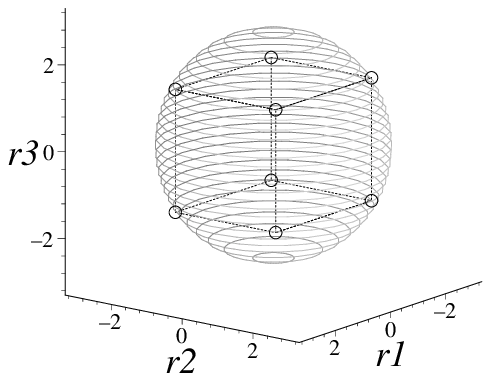} &
\includegraphics{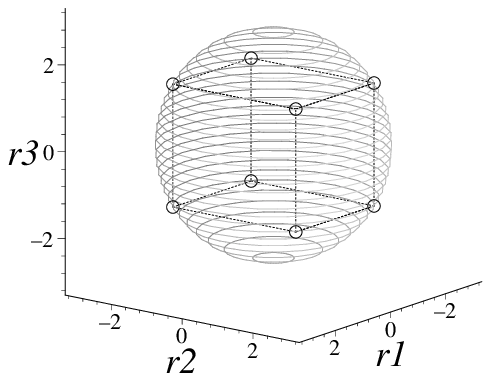} &
\includegraphics{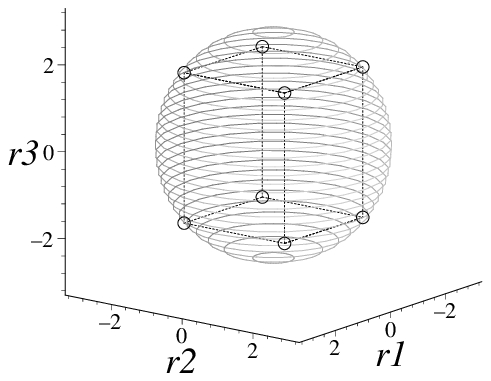} \\
$|p;1,1,0\rangle $ & $|p;1,0,1\rangle $ & $|p;0,1,1\rangle $ \\[3mm]
\includegraphics{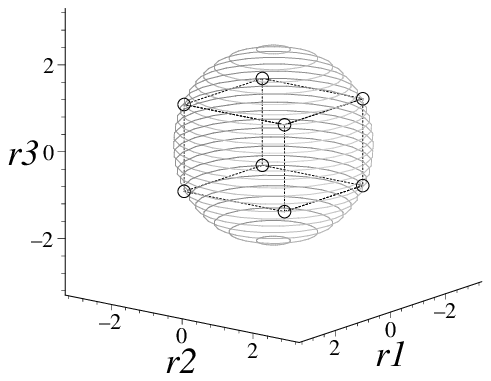} &
\includegraphics{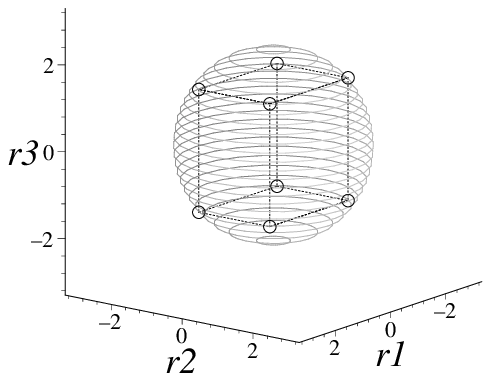} &
\includegraphics{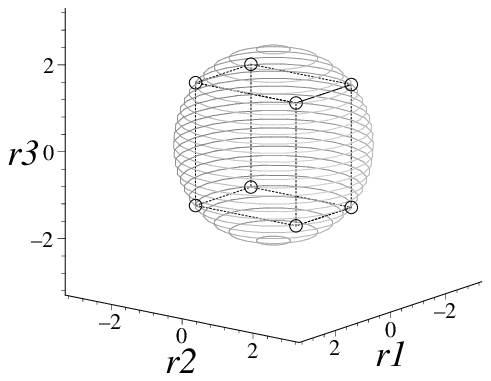} \\
 & $|p;1,1,1\rangle $ & \\[3mm]
 & \includegraphics{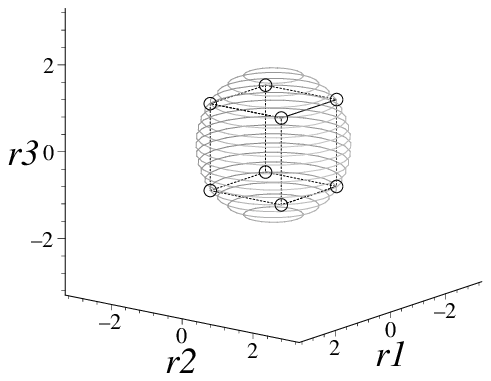} &  \\
\end{tabular}
\end{center}
\noindent Figure 1. Identification of the results of all possible measurements of
the coordinates of the particle for each of the stationary states
$|p;\T\ra$ of $W(p)$ with $p>2$.
The eight small circles on each sphere are the nests, 
the places where the oscillating particle can be spotted.
The nests in the states of the first ($q=0$), the second ($q=1$), 
the third ($q=2$) and the last line ($q=3$) from the top, are on spheres with radii $\varrho_0>\varrho_1>\varrho_2>\varrho_3$, see (\ref{3.7}), and energies
$E_0>E_1>E_2>E_3$, (see \ref{3.1}), respectively.

\newpage
\begin{center}
\begin{tabular}{ccc}
 & $|2;0,0,0\rangle$ & \\[3mm]
 & \includegraphics{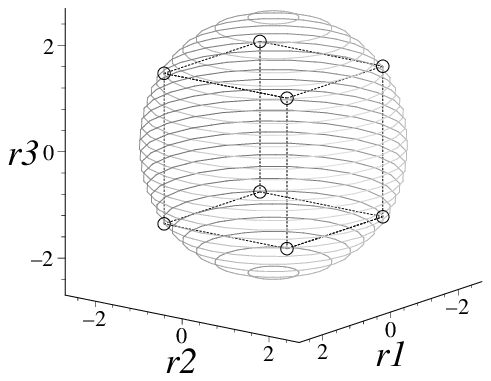} &  \\
$|2;1,0,0\rangle $ & $|2;0,1,0\rangle $ & $|2;0,0,1\rangle $\\[3mm]
\includegraphics{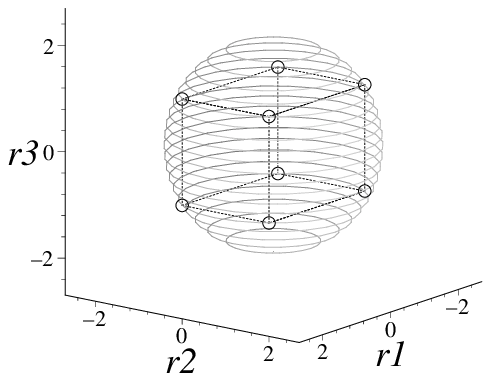} &
\includegraphics{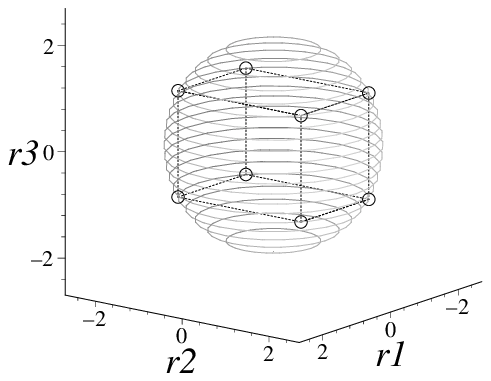} &
\includegraphics{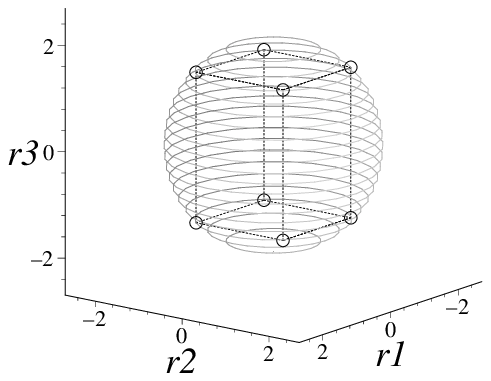} \\
$|2;1,1,0\rangle $ & $|2;1,0,1\rangle $ & $|2;0,1,1\rangle $ \\[3mm]
\includegraphics{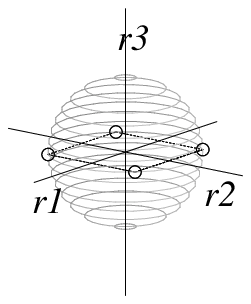} &
\includegraphics{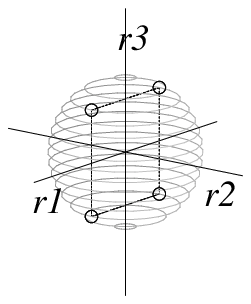} &
\includegraphics{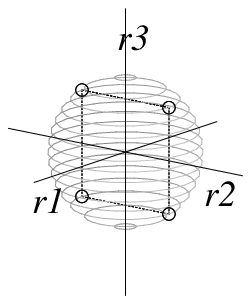} \\
\end{tabular}
\end{center}
\noindent Figure 2. 
Identification of the results of all possible measurements of the
coordinates of the particle for each of the stationary states
$|p;\T\ra$ of $W(p)$  with $p=2$.  
For $q=0$ or $q=1$ (the states on the first and the second
line from the top), the coordinates of the particle correspond to 
one of the eight nests on the sphere, indicated by circles. 
For $q=2$ (the bottom line), there are three independent lowest
energy states. For these states, there are only 4 possible nests on each sphere with
radius $\sqrt{2}$. These 4 nests are in the $r_1r_2$-plane for $\T=(1,1,0)$, in the
$r_1r_3$-plane for $\T=(1,0,1)$, and in the $r_2r_3$-plane for $\T=(0,1,1)$.

\newpage
\noindent 
\begin{center}
\begin{tabular}{ccc}
 & $|1;0,0,0\rangle$ & \\[3mm]
 & \includegraphics{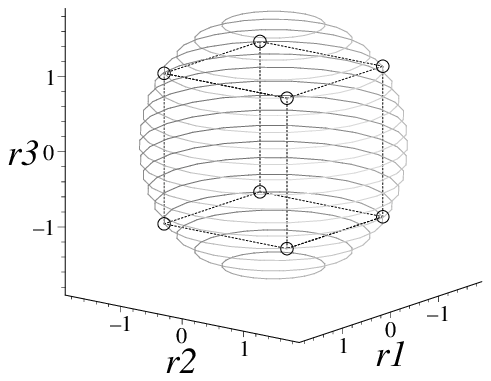} &  \\
$|1;1,0,0\rangle $ & $|1;0,1,0\rangle $ & $|1;0,0,1\rangle $\\[3mm]
\includegraphics{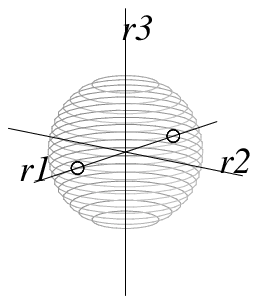} &
\includegraphics{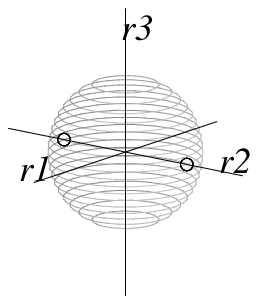} &
\includegraphics{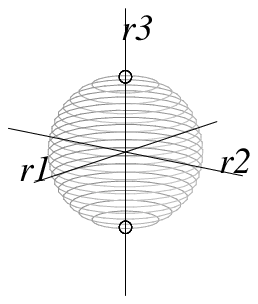} \\
\end{tabular}
\end{center}
\noindent Figure 3. 
Identification of the results of all possible measurements of the
coordinates of the particle for each of the stationary states
$|p;\T\ra$ of $W(p)$ with $p=1$.
For $q=0$ (the top picture) the coordinates of the particle
corresponds to one of the eight nests on the sphere, indicated by circles. For $q=1$
(the bottom line), there are  three independent lowest energy states. For each
such state the nests are at opposite poles on a sphere with radius $1$. 

\end{document}